# Charged Higgs Contribution on $B_c \to (D_s, D_s^*)l^+l^-$

**P. Maji\*, S. Biswas, P. Nayek and S. Sahoo\*\***

Department of Physics, National Institute of Technology Durgapur

Durgapur-713209, West Bengal, India

\*E-mail: majipriya@gmail.com, \*\*E-mail: sukadevsahoo@yahoo.com

**Abstract**

In this paper, we study rare semileptonic decays of $B_c$ meson in the context of type-I, II, and III two Higgs doublet model. We follow the relativistic quark model for parameterizing the form factors used in matrix elements of weak transitions between the corresponding meson states. We investigate the observables such as branching ratio, lepton polarization asymmetry, forward-backward asymmetry, etc. and analyze the dependence of these quantities on the model parameters. We have found that there are noticeable sensitivity of these observables for charged Higgs boson, which may provide a powerful probe to the standard model and new physics beyond it.

Keywords: Standard model, exclusive rare decays, models beyond the standard model, extensions of electroweak Higgs sector.

## I. Introduction

Motivated by the anomalies present in $b \to sl^+l^-$ neutral current decays, the rare B-meson decays induced by $b \to s$ flavor-changing neutral current (FCNC) transitions become the main attraction of researchers. These channels are forbidden at tree level but appear at the loop level within the standard model (SM). Rare B decays offer the possibility of more precise determination of the SM parameters and are very sensitive to the new intermediate particles. Within the SM background, most of the experimental observations could be explained except some of the discrepancies lying in the angular observable $P_5'$ [1] of $B \to K^*\mu^+\mu^-$, branching ratio of $B \to \phi\mu^+\mu^-$ [2], lepton flavor non-universality parameter $R_{K^{(*)}}$ [3,4] and $R_{D^{(*)}}$ [5] etc. With the possibility of upcoming production of a large number of $B_c$ meson at future LHC run (with the luminosity values of $\mathcal{L} = 10^{34}$cm$^{-2}$s$^{-1}$ and $\sqrt{s} = 14$TeV), one might explore the rare semileptonic $B_c$ decays to $D_s^{(*)}l^+l^-$ induced by FCNC $b \to s$ transitions. These transitions are useful probes to test the SM, and various extensions of the SM [6, 7] due to the following reasons: (i) Analysis of such transitions provide valuable information about the nature of pseudo-scalar $D_s$ meson and vector $D_s^*$ meson, along with the strong interactions inside them. (ii) The polarization asymmetries, CP, and T violations could be explored from the form factors of these transitions. (iii) These could offer a new framework to calculate small CKM elements $V_{tq}(q = s, b)$ and leptonic decay constants of $B_c$ and $D_s$ meson with high accuracy. (iv) The possible involvement of the fourth generation, SUSY particle [8], and light dark matter [9] might be present in the loop transitions of $b \to s$.

The $B_c$ meson [10] is the only one particle composed of two heavy quarks $b$ and $c$, which are of different charge and flavor. Those heavy quarks are bound to the lowest state to form $B_c$ meson and thus several properties of its decay modes are different from other flavor neutral processes. Since other excited states of $\bar{b}c$ lie below the threshold of decay into the pair of $B$ and $D$ mesons, such states decay weakly without annihilation due to electromagnetic and strong interactions. It is expected to be an ideal system for the study of weak decays of heavy quarks. Either of the heavy quarks ($b$ or $c$) can decay individually unlike $B_{u,d}$ or $B_s$ meson. As there is only one hadronic final product, among numerous decay channels semileptonic $B_c$ meson decays are relatively clean in the theoretical aspect. Further study of the rare weak $B_c$ decays is based on the electroweak effective Hamiltonian. The QCD corrections to these modes due to hard gluon exchanges are likely to have more importance and resum of large logarithms is needed through renormalization group methods. For the investigation of the exclusive rare decays, non-perturbative methods are required to calculate the hadronic matrix elements between initial and final meson states, which are typically parameterized in terms of covariant form factors. Apparently, such calculations are model dependent. Previously the semileptonic $B_c$ decays have been analyzed in several approaches. In ref. [11], authors described a detailed study of the exclusive semileptonic $B_c$ decays in the framework of Bauer-Stech-Wirbel. In refs. [12-14], the studies were done in the relativistic and/or constituent quark model, whereas in refs. [15, 16], $B_c \to D_s^* l^+ l^-$ channels have been investigated in the SM with the fourth generation and supersymmetric models. In refs. [17, 18], the authors presented the three-point QCD approach for their analysis. The light-front quark model was adopted by the authors in refs. [19-20] for their needful probes. In ref. [21] authors explored the perturbative QCD approach to study the semileptonic $B_c$ decay channels. New physics contributions to $B_c \to D_s^* l^+ l^-$ decay have been studied extensively in $Z'$ model [22], single universal extra dimension model [23] and also analyzed in model-independent way using effective Hamiltonian approach [24, 25].

In this paper, we study the rare $B_c \to (D_s, D_s^*) \mu^+ \mu^-$ decay adopting the background of QCD-motivated relativistic quark model [14] and supplement the previous analysis by considering the effect of charged Higgs boson in the two Higgs doublet model (2HDM) [26-28]. It is one of the well-motivated and useful extensions of the SM. Having two complex scalar doublets $\Phi_1$ and $\Phi_2$ FCNC at tree level transition is allowed in 2HDM, but this could be avoided by providing an ad hoc discrete symmetry. The possible way out to avoid FCNC which implies retaining flavor conservation at tree level is to couple all the quarks to $\Phi_2$ instead of $\Phi_1$, which is mostly known as type I. The second option somehow harmonizes with the minimal supersymmetric model (MSSM), i.e., to couple the down-type quarks only to $\Phi_1$ whereas the up-type quarks couple to $\Phi_2$, which is commonly known to be type II [29, 30]. Consequently, there are several works done in search of a generic 2HDM without discrete symmetries as in types I and II, known to be type III. In this type, all the quarks couple to both the doublets $\Phi_1$ and $\Phi_2$ providing allowed FCNC at tree level [31, 32]. It indicates that type III should be parameterized such that the tree-level FCNC for the first two generations are suppressed whereas FCNC for the third generation might be non-zero until they violate the experimental data like $B - \bar{B}$ mixing, $Br(B_s \to \mu^+ \mu^-)$, $b \to s\gamma$, etc.

Our paper is arranged as follows. In sec. II, we describe the general structure of 2HDM. The next section contains the formulation of the effective weak Hamiltonian for $b \to s l^+ l^-$ transitions in the SM and 2HDM. We also discuss the hadronic matrix elements of weak current between meson states for $B_c \to D_s$ and $B_c \to D_s^*$ channels and their parameterization in terms of various meson to meson form factors. In sec. IV, helicity amplitudes for $B_c \to D_s \mu^+\mu^-$ and $B_c \to D_s^* \mu^+\mu^-$ decay modes are presented. This section also contains the descriptions of other physical observable like decay rate, lepton forward-backward asymmetry, polarization fraction, lepton polarization asymmetry and lepton flavor universality (LFU) parameter in 2HDM. Sec. V is devoted to the numerical calculation and graphical representation of these observables. Sec. VI contains a summary of the whole study with concluding remarks.

## II. The general structure of 2HDM

Before starting the detail discussion about the effective Hamiltonian, we would give a brief description of some essential points of the 2HDM. As we have already discussed before that 2HDM has two complex scalar Higgs doublets, unlike the SM, which has only one. Type I, II and III have been also described before. Model II has gained more attraction due to the same nature of the Higgs sector as of the supersymmetric models. The Higgs sector contains three neutral Higgs bosons $H^0$, $h^0$ and $A^0$, and a pair of charged Higgs bosons $H^\pm$. The interaction vertices of the Higgs boson and fermions of these models are dependent on the ratio $\tan\beta = \frac{v_2}{v_1}$, where $v_1$ and $v_2$ are the vacuum expectation values (VEVs) of the first and second Higgs doublets, respectively. $\tan\beta$ is a free parameter of the model I and II, which usually obtains constraints from $B - \bar{B}$ mixing, $K - \bar{K}$ mixing, $b \to s\gamma$ decay width, and semileptonic decay $b \to c\tau\bar{\nu}$ which are previously defined as [33, 34]:

$$0.8 \leq \tan\beta \leq 0.6 \left(\frac{m_{H^+}}{1\text{GeV}}\right).$$

Recently in ref. [35], these models with softly broken $\mathbb{Z}_2$ symmetry have considered another prior for statistical analysis in a Bayesian fit as follows:

$-1.1 \leq \log(\tan\beta) \leq 1.7$ (equivalent to $0.08 \leq \tan\beta \leq 50$) [29],
$130 \text{ GeV} \leq m_{H^0}, m_A, m_{H^+} \leq 1.6 \text{ TeV}$.

Type III 2HDM is the most general one among all the three types of this model. Since other popular models like type I and II are special cases of type III so here we will present the details of the type III model only.

The vacuum structure of 2HDM is generally very rich and the scalar potential contains 14 parameters, which can have CP-conserving, CP-violating, and charge-violating minima. Here we assume CP is conserved in the Higgs sector and not spontaneously broken which means CP-odd Higgs state is not mixed with CP-even Higgs state. Under these assumption the most general form of scalar potential for two doublets $\Phi_1$ and $\Phi_2$ with hypercharge +1 is,

$$V = m_{11}^2 \phi_1^\dagger \phi_1 + m_{22}^2 \phi_2^\dagger \phi_2 - m_{12}^2 (\phi_1^\dagger \phi_2 + \phi_2^\dagger \phi_1) + \frac{\lambda_1}{2}(\phi_1^\dagger \phi_1)^2 + \frac{\lambda_2}{2}(\phi_2^\dagger \phi_2)^2$$
$$+ \lambda_3 \phi_1^\dagger \phi_1 \phi_2^\dagger \phi_2 + \lambda_4 \phi_1^\dagger \phi_2 \phi_2^\dagger \phi_1 + \frac{\lambda_5}{2}\left[(\phi_1^\dagger \phi_2)^2 + (\phi_2^\dagger \phi_1)^2\right] \quad (1)$$

The bilinear terms proportional to $m_{12}^2$ softly break $\mathbb{Z}_2$ symmetry. The minimization of this potential gives

$$\langle\Phi_1\rangle_0 = \begin{pmatrix} 0 \\ \frac{v_1}{\sqrt{2}} \end{pmatrix}, \langle\Phi_2\rangle_0 = \begin{pmatrix} 0 \\ \frac{v_2}{\sqrt{2}} \end{pmatrix}. \tag{2}$$

There are two SU(2) scalar doublets $\Phi_a(a=1,2)$ which are parameterized as

$$\Phi_a = \begin{pmatrix} \phi_a^+ \\ \frac{1}{\sqrt{2}}[v_a + \rho_a + i\eta_a] \end{pmatrix}, \tag{3}$$

where $v_1$ and $v_2$ satisfy $v^{SM} = \sqrt{v_1^2 + v_2^2}$ and $v^{SM} = 246.22$ GeV which is vacuum expectation value (VEV) of SM [36]. With these two doublets one can obtain eight fields three of which are Goldstone bosons ($G^\pm$, $G^0$) and get eaten by $W^\pm$ and $Z^0$ gauge bosons for their mass generation. The remaining five are physical scalar fields giving rise to two CP-even neutral scalars ($h^0, H^0$), one CP-odd pseudoscalar ($A^0$) and one pair of charged scalar ($H^\pm$). These fields could be defined as

$$\begin{pmatrix} \phi_1^+ \\ \phi_2^+ \end{pmatrix} = \begin{pmatrix} \cos\beta & -\sin\beta \\ \sin\beta & \cos\beta \end{pmatrix} \begin{pmatrix} G^+ \\ H^+ \end{pmatrix},$$

$$\begin{pmatrix} \eta_1 \\ \eta_2 \end{pmatrix} = \begin{pmatrix} \cos\beta & -\sin\beta \\ \sin\beta & \cos\beta \end{pmatrix} \begin{pmatrix} G^0 \\ A^0 \end{pmatrix},$$

and

$$\begin{pmatrix} \rho_1 \\ \rho_2 \end{pmatrix} = \begin{pmatrix} \cos\alpha & -\sin\alpha \\ \sin\alpha & \cos\alpha \end{pmatrix} \begin{pmatrix} H^0 \\ h^0 \end{pmatrix}. \tag{4}$$

The mixing angles $\alpha$ and $\beta$ satisfy

$$\tan\beta = \frac{v_2}{v_1}$$

and

$$\tan\alpha = \frac{2(-m_{12}^2 + \lambda_{345}v_1v_2)}{m_{12}^2(v_2/v_1 - v_1/v_2) + \lambda_1 v_1^2 - \lambda_2 v_2^2} \tag{5}$$

with $\lambda_{345} \equiv \lambda_3 + \lambda_4 + \lambda_5$. The two parameters $\alpha$ and $\beta$ determine the interactions of the various Higgs fields with the vector bosons and fermions. The masses of the physical scalars can be written in terms of parameters which appear in the expression of potential as [37]

$$m_{H^0}^2 = M^2 \sin^2(\alpha-\beta) + (\lambda_1 \cos^2\alpha \cos^2\beta + \lambda_2 \sin^2\alpha \sin^2\beta + \frac{\lambda_{345}}{2}\sin 2\alpha \sin 2\beta)v^2$$

$$m_{h^0}^2 = M^2 \cos^2(\alpha-\beta) + (\lambda_1 \sin^2\alpha \cos^2\beta + \lambda_2 \cos^2\alpha \sin^2\beta - \frac{\lambda_{345}}{2}\sin 2\alpha \sin 2\beta)v^2$$

$$m_{A^0}^2 = M^2 - \lambda_5 v^2$$

$$m_{H^\pm}^2 = M^2 - \frac{\lambda_4 + \lambda_5}{2}v^2 \tag{6}$$

where $M^2 = m_{12}^2/\sin\beta\cos\beta$ is the $\mathbb{Z}_2$ breaking parameter.

The stability and unitarity constraints on quartic couplings of general 2HDM are as follows

i) Stability: The quartic couplings should agree to the following relations for the scalar potential to be bounded [38] from below

$$\lambda_{1,2} > 0, \lambda_3 > -(\lambda_1\lambda_2)^{1/2} \text{ and } \lambda_3 + \lambda_4 - |\lambda_5| > -(\lambda_1\lambda_2)^{\frac{1}{2}} \tag{7}$$

Moreover, the stability of the electroweak vacuum implies that

$$m_{11}^2 + \frac{\lambda_1 v_1^2}{2} + \frac{\lambda_3 v_2^2}{2} = \frac{v_2}{v_1}[m_{12}^2 - (\lambda_4 + \lambda_5)\frac{v_1 v_2}{2}]$$

$$m_{22}^2 + \frac{\lambda_2 v_2^2}{2} + \frac{\lambda_3 v_1^2}{2} = \frac{v_1}{v_2}[m_{12}^2 - (\lambda_4 + \lambda_5)\frac{v_1 v_2}{2}] \tag{8}$$

which allows to express $m_{11}^2$ and $m_{22}^2$ in terms of $\mathbb{Z}_2$ breaking term $m_{12}^2$ and the quartic coupling $\lambda_{1-5}$. Combining these constraints with necessary and sufficient condition one can develop a global minima at $(v_1, v_2)$ which is [39]

$$m_{12}^2(m_{11}^2 - m_{22}^2\sqrt{\lambda_1/\lambda_2})(\tan\beta - \sqrt[4]{\lambda_1/\lambda_2}) > 0. \tag{9}$$

ii) Perturbative unitarity: The unitarity condition of the S-wave component of the scalar amplitude constrain the spectrum of scalars in 2HDM. That condition denotes following inequalities [40, 41],

$$|a_\pm|, |b_\pm|, |c_\pm|, |d_\pm|, |e_{1,2}|, |f_1|, |p_1| < 8\pi \tag{10}$$

where

$$a_\pm = \frac{3}{2}(\lambda_1 + \lambda_2) \pm \sqrt{\frac{9}{4}(\lambda_1 - \lambda_2)^2 + (2\lambda_3 + \lambda_4)^2}$$

$$b_\pm = \frac{1}{2}(\lambda_1 + \lambda_2) \pm \frac{1}{2}\sqrt{(\lambda_1 - \lambda_2)^2 + 4\lambda_4^2}$$

$$c_\pm = \frac{1}{2}(\lambda_1 + \lambda_2) \pm \frac{1}{2}\sqrt{(\lambda_1 - \lambda_2)^2 + 4\lambda_5^2}$$

$$d_+ = \lambda_3 + 2\lambda_4 + 3\lambda_5, d_- = \lambda_3 + \lambda_5$$

$$e_1 = \lambda_3 + 2\lambda_4 - 3\lambda_5, e_2 = \lambda_3 - \lambda_5$$

$$f_1 = \lambda_3 + \lambda_4, p_1 = \lambda_3 - \lambda_4. \tag{11}$$

The additional scalar of beyond SM contributes to the gauge boson vacuum polarization and this leads to constraint all three 2HDM scenarios with electroweak precision data. When the mass of extra scalar bosons are large compared to $m_Z$, their contributions could be represented in terms of three oblique parameters $S, T,$ and $U$. Among these three particularly the T parameter bounds the mass splitting between $M_{H^0}$ and $M_{H^\pm}$ and is sensitive to weak isospin violation in the scenario where $h^0$ is identified as SM-like Higgs. The $S$ $(S+U)$ parameter implies NP contributions to neutral (charged) current processes at different energy scales. The third parameter U is constrained only by the $W$ boson mass and decay width. The explicit expressions of $S, T$ and $U$ could be found in ref. [42] and their updated values are as follows

$$S = 0.04 \pm 0.11, T = 0.09 \pm 0.14, U = -0.02 \pm 0.11,$$

with correlation coefficients of $+0.92$ between $S$ and $T$, $-0.68$ between $S$ and $U$, $-0.87$ between $T$ and $U$. Fixing $U = 0$ one obtains $S|_{u=0} = 0.04 \pm 0.08$ and $T|_{u=0} = 0.08 \pm 0.07$. Except for the masses of scalars, all other parameters of 2HDM are free to vary. These oblique parameters constrain the values of $M_{H^0}$ and $M_{A^0}$ to be close to $M_{H^\pm}$ for fixed $\beta - \alpha = \pi/2$. Taking out this restriction would relax the constraint to have either $M_A$ close to $M_{H^\pm}$ or $M_{H^0}$ larger than $M_{H^\pm}$.

In general framework of type III 2HDM, the FCNC parameters have been considered to be small at low energy like $E \sim E_{ew}$. They may increase and could have large positive contributions when one-loop renormalization group equations (RGEs) running up the energy scale. They behave differently in the FCNC Yukawa sector at higher energy than at the low energy (electroweak). The Landau pole is such an energy where the couplings blow up and become infinite. Any finite value of coupling would have Landau pole at some energy level above which new physics particles tend to exist. In our analysis, we have considered the most general scenario which is based on the assumption that $m_{h^0} = 125$ GeV. This light Higgs state is expected to be the SM-like Higgs boson. We also impose the lower bounds as $m_{H^0}, m_{A^0} > m_{h^0}$ whereas for charged scalar the bound is fixed at $m_{H^\pm} \geq 160$ GeV. These bounds would induce Landau pole at an intermediate scale while running the RGE at higher energies. The Cheng-Sher (CS) ansatz [43,44] allow to construct one loop RGEs for evaluating the Yukawa parameters as well as the gauge and scalar quartic couplings at low energy scale $E \sim E_{ew}$ to fulfill the FCNC suppression restriction.

As discussed earlier, type III 2HDM allows the coupling of up-type and down-type quark with both the doublets, so without losing the generality a new basis could be considered where the first Higgs doublet is responsible for the mass generation to all gauge bosons and fermions within the SM and new Higgs field is produced from second doublet. This fact promotes to formulate of the flavor changing (FC) part of Yukawa Lagrangian at tree level as

$$\mathcal{L}_{Y,FC}^{III} = \xi_{ij}^U \bar{Q}_{iL} \tilde{\phi}_2 U_{jR} + \xi_{ij}^D \bar{Q}_{iL} \phi_2 D_{jR} + \xi_{ij}^l \bar{l}_{iL} \phi_2 l_{jR} + h.c., \quad (12)$$

where, $i, j$ represent generation indices, $\bar{Q}_{iL}$ is the left-handed fermion doublet, $U_{jR}$ and $D_{jR}$ are right-handed quark singlets, $\bar{l}_{iL}$ and $l_{jR}$ are left-handed SU(2) lepton doublet and SU(2) singlet respectively, $\tilde{\phi}_2 = i\sigma_2 \phi_2$. Here, $\xi_{ij}^{U,D,l}$ are generally non-diagonal coupling matrices. The couplings $\xi^{U,D}$ are the open window for tree-level FCNCs and can be parameterized for FC charged interactions as

$$\xi_{ch}^U = \xi_{neutral} V_{CKM}, \xi_{ch}^D = V_{CKM} \xi_{neutral},$$

All the states above are weak states and when we do a proper rotation and diagonalization of the mass matrices for fermions and Higgses, eq. (12) could be rewritten in terms of mass eigenstates as follows [45]

$$\mathcal{L}_{Y,FC}^{III} = \frac{1}{\sqrt{2}}[\bar{U}_i\hat{\xi}_{ij}^U U_j + \bar{D}_i\hat{\xi}_{ij}^D D_j + \bar{l}_i\hat{\xi}_{ij}^l l_j]H^0 - \frac{i}{\sqrt{2}}[\bar{U}_i\gamma_5\hat{\xi}_{ij}^U U_j + \bar{D}_i\gamma_5\hat{\xi}_{ij}^D D_j + \bar{l}_i\gamma_5\hat{\xi}_{ij}^l l_j]A^0$$
$$+ \bar{U}_i[\hat{\xi}_{ij}^U V_{CKM} P_L - V_{CKM}\hat{\xi}_{ij}^D P_R]D_j H^{\pm} + h.c., \quad (13)$$

where, $U_i, D_i, l_i$ are mass eigenstates of up- and down-type quarks and leptons, respectively. $V_{CKM} = (V_L^U)^\dagger V_L^D$ is the usual CKM matrix element, $P_{L(R)} = \frac{(1\mp\gamma_5)}{2}$ are the projection operators and $\hat{\xi}_{ij}^{U,D,l}$ are FC Yukawa quark matrices which include FCNC couplings. Continuing our discussion further we adopt the Cheng-Sher (CS) ansatz in this work [45],

$$\hat{\xi}_{ij}^{U,D,l} = \lambda_{ij}\frac{\sqrt{2m_i m_j}}{v^{SM}}, \quad (14)$$

This ansatz highly secures the suppression of FCNC within the first two generations for small quark masses, while larger freedom is allowed for third-generation FCNC. The interesting fact about that complex nature of $\lambda_{ij}$'s is that it includes electric dipole moments of electrons and quarks as a consequence of the explicit CP violation due to the complex phase in the charged Higgs sector. So, for simplicity we consider $\hat{\xi}^{U,D}$ to be diagonal and consequently $\lambda_{ij}$'s are also diagonal.

To have a truthful type III 2HDM one would expect $\lambda_{ij}\sim 1$, but this is a fairly loose requirement since there are unknown mixing angles. Using Cheng-Sher ansatz and assuming $m_{A^0} = 120$ GeV, Branco et al. [46] found the couplings as $(\lambda_{ds}, \lambda_{uc}, \lambda_{bd}, \lambda_{bs}) \leq (0.1, 0.2, 0.06, 0.06)$. This might be challenging for type III, but increasing $m_{A^0}$ to 400 GeV all the bounds of the above couplings could be increased by more than a factor three. In [47], it is shown that the bound is increased substantially as the pseudoscalar mass increases. These bounds depend basically on the upper renormalization scale, i.e., gauge unification scale $\Lambda\sim 10^{16}$ GeV. The upper bound on the lightest CP-even scalar would be changed from 300 to about 100 GeV when $\Lambda$ is varied from $10^3$ to $10^{16}$ GeV. If we require perturbativity up to the GUT scale all the gauge and Yukawa couplings go much below $\sqrt{4\pi}$, because Yukawa couplings encounter Landau pole before the GUT scale when $\tan\beta = \frac{v_2}{v_1} = 1$. So, in order to have a phenomenologically feasible suppression of scalar mediated FCNCs in perturbative Yukawa sector, we have to consider $\lambda_{ij}\sim 1$ for $E\sim E_{ew}$. Furthermore, several bounds on other couplings are extensively reviewed in [46].

From eq. (13), we see that, at tree level $b \to sl^+l^-$ transitions have contribution due to neutral Higgs boson exchange diagrams $\sim \hat{\xi}_{bs}\hat{\xi}_{\mu\mu}\frac{1}{q^2-M_{H^0(A^0)}^2}[(\bar{b}(\gamma_5)s)(\bar{\mu}(\gamma_5)\mu)]H^0(A^0)$. **So,** it is essential to impose a natural condition to suppress such FCNCs, namely natural flavor conservation [48]. To proceed further, we consider only the charged Higgs boson contributions into account and neglect the effect of that neutral Higgs boson. Here we assume the masses of neutral Higgs bosons are heavy compared to the $b$ quark mass and thus only contribute to $b \to s\gamma$ decay. From ref. [49] we can see that neutral Higgs boson could provide a large contribution to $C_7$. At this level, we can make a choice that the couplings $\bar{\xi}_{N(neutral),is}^D (i = d, s, b)$ and $\bar{\xi}_{N(neutral),db}^D$ are negligible in order to reach the conditions $\bar{\xi}_{N,bb}^D\bar{\xi}_{N,is}^D \ll 1$ and $\bar{\xi}_{N,db}^D\bar{\xi}_{N,ds}^D \ll 1$. These choices permit us to neglect NHBs exchange diagrams [50] and $b \to sl^+l^-$ transitions receive contribution only from $W^\pm, Z, \gamma$ and charged Higgs boson. And remarkably, the charged Higgs field does not yield any new operators to $b \to sl^+l^-$ transitions, instead modify only the value of SM Wilson coefficients [51, 52].

## III. The formalism of effective Hamiltonian for $b \to sl^+l^-$ transition

The effective Hamiltonian representing $B_c \to D_s^{(*)}l^+l^-$ decays consisting $|\Delta B| = |\Delta S| = 1$ transition for both the SM and 2HDM could be expressed in terms of a set of local operators and is written in the following form [53]

$$\mathcal{H}_{eff} = -\frac{4G_F}{\sqrt{2}} V_{tb}V_{ts}^* \left\{ \sum_{i=1}^{10} C_i(\mu)O_i(\mu) + \sum_{i=1}^{10} C_{Qi}(\mu)Q_i(\mu) \right\}, \tag{15}$$

where, $G_F$ is the Fermi constant and $V_{tb}$, $V_{ts}^*$ are the corresponding CKM matrix element for $b \to s$ transition. $O_i(\mu)$ are the set of local operators and $C_i(\mu)$ are the relevant Wilson coefficients that contain short- and long-distance contributions evaluated at renormalization scale $\mu = m_b^{pole} = 4.8$ GeV. The additional operators $Q_i(\mu)$ represent the contribution coming from neutral Higgs boson (NHB) exchange diagrams, whose forms and corresponding Wilson coefficients $C_{Qi}(\mu)$ are given in ref. [54].

The new Hamiltonian responsible for the $B_c \to (D_s, D_s^*)l^+l^-$ decay mode in type III 2HDM is given as

$$\mathcal{H}_{eff}(b \to sl^+l^-) = \frac{G_F \alpha_{em}}{2\sqrt{2}\pi} V_{tb}V_{ts}^* \{C_9^{2HDM}(\mu)\bar{s}\gamma_\mu(1-\gamma_5)b(\bar{l}\gamma^\mu l)$$
$$+C_{10}^{2HDM}(\mu)\bar{s}\gamma_\mu(1-\gamma_5)b(\bar{l}\gamma^\mu\gamma_5 l) - 2m_b C_7^{2HDM}(\mu)\bar{s}i\sigma_{\mu\nu}\frac{q^\nu}{q^2}(1+\gamma_5)b(\bar{l}\gamma^\mu l)\} \tag{16}$$

where, $C_7^{2HDM}(\mu)$, $C_9^{2HDM}(\mu)$ and $C_{10}^{2HDM}(\mu)$ are as follows [54]

$$C_7^{2HDM}(\mu) = C_{7(H)}^{eff}(\mu) + |\lambda_{tt}|^2 \left( \frac{y(7-5y-8y^2)}{72(y-1)^2} + \frac{y^2(3y-2)}{12(y-1)^4}\ln y \right)$$
$$+ |\lambda_{tt}\lambda_{bb}|e^{i\theta} \left( \frac{y(3-5y)}{12(y-1)^2} + \frac{y(3y-2)}{6(y-1)^3}\ln y \right),$$

$$C_9^{2HDM}(\mu) = C_9^{eff}(\mu)$$
$$+ |\lambda_{tt}|^2 \left[ \frac{1-4\sin^2\theta_W}{\sin^2\theta_W}\frac{xy}{8}\left(\frac{1}{y-1} - \frac{1}{(y-1)^2}\ln y\right) \right.$$
$$\left. - y\left(\frac{47y^2 - 79y + 38}{108(y-1)^3} - \frac{3y^3 - 6y + 4}{18(y-1)^4}\ln y\right) \right],$$

$$C_{10}^{2HDM}(\mu) = C_{10}^{SM}(\mu)$$
$$+ |\lambda_{tt}|^2 \frac{1}{\sin^2\theta_W}\frac{xy}{8}\left(-\frac{1}{y-1} + \frac{1}{(y-1)^2}\ln y\right), \tag{17}$$

where $x = \frac{m_t^2}{m_W^2}$ and $y = \frac{m_t^2}{m_{H^\pm}^2}$ with $m_t$ and $m_W$ being the top quark and W boson mass, respectively.

For further corrections due to $b \to s\gamma$ [55], we consider other nonfactorizable effects coming from charm loop and evaluate the Wilson coefficient $C_7^{eff}$ in leading logarithm approximation as [56]

In SM, $C_{7(SM)}^{eff}(\mu) = \eta^{\frac{16}{23}}C_7(m_W) + \frac{8}{3}\left(\eta^{\frac{14}{23}} - \eta^{\frac{16}{23}}\right)C_8(m_W) + C_2(m_W)\sum_{i=1}^{8}h_i\eta^{a_i},$ (18)

In 2HDM, $C_{7(H)}^{eff}(\mu) = \eta^{\frac{16}{23}}C_7^H(m_W) + \frac{8}{3}\left(\eta^{\frac{14}{23}} - \eta^{\frac{16}{23}}\right)C_8^H(m_W) + C_2^H(m_W)\sum_{i=1}^{8}h_i\eta^{a_i},$ (19)

where, $C_2(m_W) = 1$, $C_2^H(m_W) = 0$ and $C_7(m_W), C_7^H(m_W), C_8(m_W), C_8^H(m_W)$ are given in ref. [57]. The coefficients $a_i$ and $h_i$ are given as [58, 59],

$$\begin{aligned}a_i &= (14/23, \quad 16/23, \quad 6/23, \quad -12/23, \quad 0.4086, \quad -0.4230, \quad -0.8994, \quad 0.1456), \\ h_i &= (2.2996, \quad -1.0880, \quad -3/7, \quad -1/14, \quad -0.6494, \quad -0.0380, \quad -0.0186, \quad -0.0057).\end{aligned}$$

The parameter $\eta$ in eq. (18, 19) is defined as, $\eta = \frac{\alpha_s(\mu_W)}{\alpha_s(\mu_b)}$.

The Wilson coefficient $C_9^{eff}$ in the NDR scheme could be found in refs. [57, 58]. $C_9^{eff}$ contains the contribution from perturbative terms, i.e., known as short-distance contribution. It also contains charm loop contribution arising due to two lowest resonances $J/\psi(1S)$ and $\psi(2S)$ [14], which could be obtained by a non-perturbative approach. At $\mu_b$ scale, for the coefficient $C_j$ ($j = 1 \ldots 6$) we have $C_j = \sum_{i=1}^{6} k_{ji}\eta^{a_i}$. For further details, one can see ref. [60].

In type III 2HDM, mass of charged Higgs boson and the coefficients $\lambda_{tt}, \lambda_{bb}$ are free parameters among which $\lambda_{tt}$ and $\lambda_{bb}$ are complex in nature. Here, $\lambda_{tt}\lambda_{bb} = |\lambda_{tt}\lambda_{bb}|e^{i\theta}$ where $\theta$ represents the only single CP phase of the vacuum. Thus $\lambda_{ij}$ allows the charged Higgs boson to produce constructive or destructive interference with the SM contribution.

One important point is that we can obtain model I and II from model III by substituting the following relations:

$\lambda_{tt} = \cot\beta$, $\lambda_{tt}\lambda_{bb} = -\cot^2\beta$ for type I
$\lambda_{tt} = \cot\beta$, $\lambda_{tt}\lambda_{bb} = 1$ for type II

The hadronic matrix elements for $B_c \to D_s\mu^+\mu^-$ decays could be written in terms of three invariant meson to meson transition form factors. These are

$$\begin{aligned}\langle D_s|\bar{s}\gamma^\mu b|B_c\rangle &= f_+(q^2)\left[p_{B_c}^\mu + p_{D_s}^\mu - \frac{M_{B_c}^2 - M_{D_s}^2}{q^2}q^\mu\right] + f_0(q^2)\frac{M_{B_c}^2 - M_{D_s}^2}{q^2}q^\mu, \\ \langle D_s|\bar{s}\sigma^{\mu\nu}q_\nu b|B_c\rangle &= \frac{if_T(q^2)}{M_{B_c}+M_{D_s}}\left[q^2(p_{B_c}^\mu + p_{D_s}^\mu) - (M_{B_c}^2 - M_{D_s}^2)q^\mu\right].\end{aligned}$$ (20)

Similarly, for $B_c \to D_s^*\mu^+\mu^-$ channels the hadronic matrix elements could be parameterized in terms of seven invariant form factors. These are

$$\langle D_s^*|\bar{s}\gamma^\mu b|B_c\rangle = \frac{2iV(q^2)}{M_{B_c} + M_{D_s^*}}\epsilon^{\mu\nu\rho\sigma}\epsilon_\nu^* p_{B_c\rho}p_{D_s^*\sigma},$$

$$\langle D_s^*|\bar{s}\gamma^\mu\gamma_5 b|B_c\rangle = 2M_{D_s^*}A_0(q^2)\frac{\epsilon^*.q}{q^2}q^\mu + (M_{B_c} + M_{D_s^*})A_1(q^2)\left(\epsilon^{*\mu} - \frac{\epsilon^*.q}{q^2}q^\mu\right)$$
$$- A_2(q^2)\frac{\epsilon^*.q}{(M_{B_c} + M_{D_s^*})}\left[p_{B_c}^\mu + p_{D_s^*}^\mu - \frac{M_{B_c}^2 - M_{D_s^*}^2}{q^2}q^\mu\right],$$

$$\langle D_s^*|\bar{s}\sigma^{\mu\nu}q_\nu b|B_c\rangle = 2T_1(q^2)\epsilon^{\mu\nu\rho\sigma}\epsilon_\nu^* p_{B_c\rho}p_{D_s^*\sigma},$$

$$\langle D_s^*|\bar{s}\sigma^{\mu\nu}\gamma_5 q_\nu b|B_c\rangle = T_2(q^2)[(M_{B_c}^2 - M_{D_s^*}^2)\epsilon^{*\mu} - (\epsilon^*.q)(p_{B_c}^\mu + p_{D_s^*}^\mu)]$$

$$+T_3(q^2)(\epsilon^*\cdot q)[q^\mu - \frac{q^2}{M_{B_c}^2 - M_{D_s^*}^2}(p_{B_c}^\mu + p_{D_s^*}^\mu)], \tag{21}$$

where, $q^\mu = (p_B - p_{D_s}, p_{D_s^*})^\mu$ is the four-momentum transfer and $\epsilon_\mu$ is polarization vector of $D_s^*$ meson. For evaluating the form factors, we adopt the relativistic quark model [14] approach.

The Feynman diagrams for corresponding quark transitions are shown below in figs. 1 and 2.

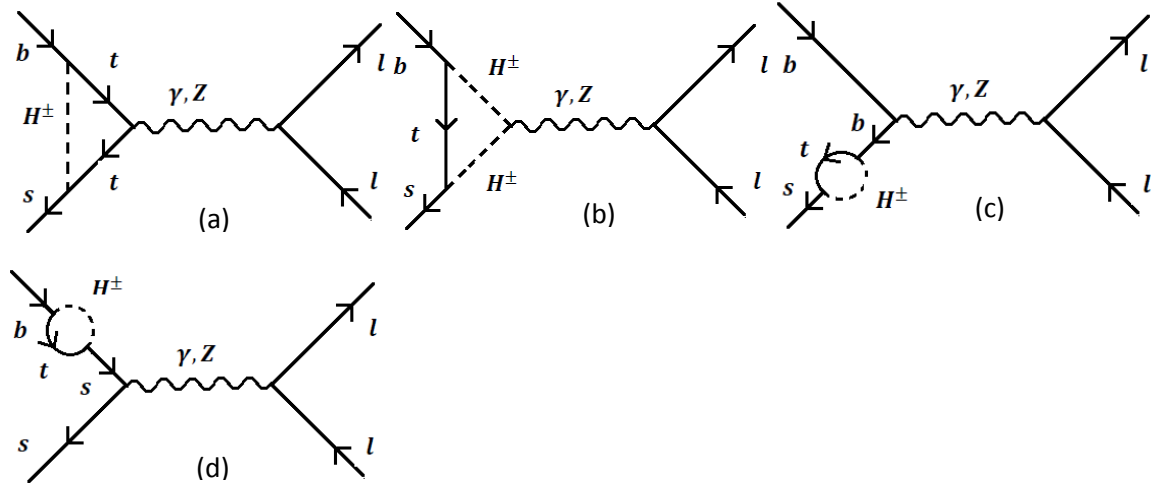

Fig. 1: Photon and Z penguin diagrams with charged Higgs boson contribution.

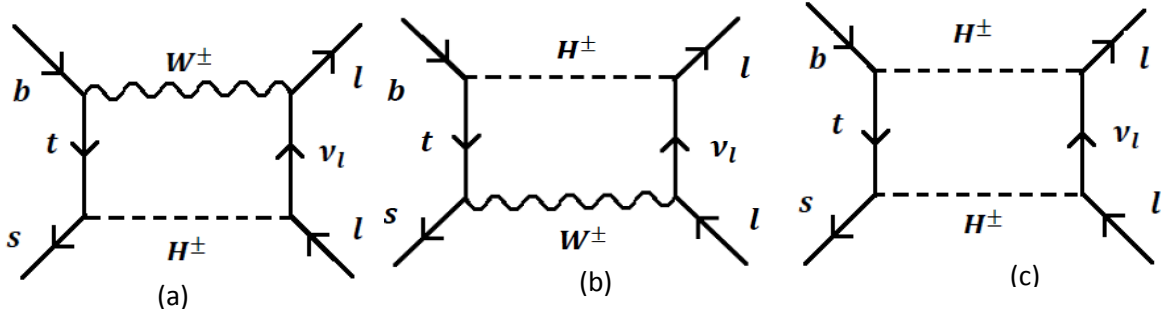

Fig. 2: W box diagrams with charged Higgs boson contribution.

## IV. Helicity Amplitudes and Physical Observables

For obtaining helicity amplitudes we consider the pattern after refs. [14, 15] and reuse the techniques of refs. [61, 62]. Our helicity amplitudes for $B_c \to D_s l^+ l^-$ decay mode are written as

$$H_\pm^{(i)} = 0,$$

$$H_0^{(1)} = \sqrt{\frac{\lambda}{q^2}}\left[C_9^{2HDM} f_+(q^2) + C_7^{2HDM}\frac{2m_b}{M_{B_c} + M_{D_s}} f_T(q^2)\right],$$

$$H_0^{(2)} = \sqrt{\frac{\lambda}{q^2}} C_{10}^{2HDM} f_+(q^2),$$

$$H_t^{(1)} = \frac{M_{B_c}^2 - M_{D_s}^2}{\sqrt{q^2}} C_9^{2HDM} f_0(q^2),$$

$$H_t^{(2)} = \frac{M_{B_c}^2 - M_{D_s}^2}{\sqrt{q^2}} C_{10}^{2HDM} f_0(q^2). \tag{22}$$

As the final meson $D_s$ is a pseudo-scalar meson and does not have any polarization direction, so the transverse helicity amplitudes for $B_c \to D_s l^+ l^-$ channel are 0.

Similarly for $B_c \to D_s^* l^+ l^-$ modes, the hadronic helicity amplitudes are

$$H_\pm^{(1)} = -(M_{B_c}^2 - M_{D_s^*}^2)\left[C_9^{2HDM}\frac{A_1(q^2)}{(M_{B_c} - M_{D_s^*})} + \frac{2m_b}{q^2}C_7^{2HDM}T_2(q^2)\right]$$
$$\pm \sqrt{\lambda}\left[C_9^{2HDM}\frac{V(q^2)}{(M_{B_c} + M_{D_s^*})} + \frac{2m_b}{q^2}C_7^{2HDM}T_1(q^2)\right],$$

$$H_\pm^{(2)} = C_{10}^{2HDM}[-(M_{B_c} + M_{D_s^*})A_1(q^2)] \pm \frac{\sqrt{\lambda}}{(M_{B_c} + M_{D_s^*})}C_{10}^{2HDM}V(q^2),$$

$$H_0^{(1)} = -\frac{1}{2M_{D_s^*}\sqrt{q^2}}\left\{C_9^{2HDM}\left[(M_{B_c}^2 - M_{D_s^*}^2 - q^2)(M_{B_c} + M_{D_s^*})A_1(q^2) - \frac{\lambda}{M_{B_c} + M_{D_s^*}}A_2(q^2)\right]\right.$$
$$\left. + 2m_b C_7^{2HDM}\left[(M_{B_c}^2 + 3M_{D_s^*}^2 - q^2)T_2(q^2) - \frac{\lambda}{M_{B_c}^2 - M_{D_s^*}^2}T_3(q^2)\right]\right\},$$

$$H_0^{(2)} = -\frac{1}{2M_{D_s^*}\sqrt{q^2}}C_{10}^{2HDM}\left[(M_{B_c}^2 - M_{D_s^*}^2 - q^2)(M_{B_c} + M_{D_s^*})A_1(q^2) - \frac{\lambda}{M_{B_c} + M_{D_s^*}}A_2(q^2)\right],$$

$$H_t^{(1)} = \sqrt{\frac{\lambda}{q^2}} C_9^{2HDM} A_0(q^2),$$

$$H_t^{(2)} = \sqrt{\frac{\lambda}{q^2}} C_{10}^{2HDM} A_0(q^2), \tag{23}$$

where,

$$\lambda = M_{B_c}^4 + M_{D_s,D_s^*}^4 + q^4 - 2(M_{B_c}^2 M_{D_s,D_s^*}^2 + M_{D_s,D_s^*}^2 q^2 + M_{B_c}^2 q^2). \tag{24}$$

Based on the calculation of ref. [25], the three-body $B_c \to D_s l^+ l^-$ and $B_c \to D_s^* l^+ l^-$ differential decay rate could be written as,

$$\frac{d\Gamma}{dq^2} = \frac{G_F^2}{(2\pi)^3}\left(\frac{\alpha_e |V_{tb}V_{ts}^*|}{2\pi}\right)^2 \frac{\lambda^{1/2}q^2}{48 M_{B_c}^3}\sqrt{1-\frac{4m_l^2}{q^2}}\left[H^{(1)}H^{\dagger(1)}\left(1+\frac{4m_l^2}{q^2}\right)\right.$$
$$\left. + H^{(2)}H^{\dagger(2)}\left(1-\frac{4m_l^2}{q^2}\right) + \frac{2m_l^2}{q^2}3H_t^{(2)}H_t^{\dagger(2)}\right], \tag{25}$$

where, $m_l$ is the lepton mass and

$$H^{(i)}H^{\dagger(i)} = H_+^{(i)}H_+^{\dagger(i)} + H_-^{(i)}H_-^{\dagger(i)} + H_0^{(i)}H_0^{\dagger(i)}. \tag{26}$$

The differential ratio of branching ratio also known as LFU parameter is defined as

$$R_{D_s,D_s^*}(q^2) = \frac{d\Gamma/dq^2(B_c \to (D_s, D_s^*)\mu^+\mu^-)}{d\Gamma/dq^2(B_c \to (D_s, D_s^*)e^+e^-)}. \tag{27}$$

Besides that, we also study some other observables like forward-backward asymmetry ($A_{FB}$) and the longitudinal polarization fraction ($P_L$) of the final vector meson in the decay $B_c \to D_s^* l^+ l^-$. While analyzing the channel $B \to K^* l^+ l^-$, $A_{FB}$ and $P_L$ have got wide attention both theoretically and experimentally. It is expected to collect more information on the Wilson coefficient by investigating these observables. The forward-backward asymmetry ($A_{FB}$) is given by [25]

$$A_{FB}(q^2) = \frac{3}{4}\sqrt{1 - \frac{4m_l^2}{q^2}}$$
$$\left\{\frac{Re\left(H_+^{(1)}H_+^{\dagger(2)}\right) - Re(H_-^{(1)}H_-^{\dagger(2)})}{H^{(1)}H^{\dagger(1)}\left(1 + \frac{4m_l^2}{q^2}\right) + H^{(2)}H^{\dagger(2)}\left(1 - \frac{4m_l^2}{q^2}\right) + \frac{2m_l^2}{q^2}3H_t^{(2)}H_t^{\dagger(2)}}\right\}. \tag{28}$$

A notable fact is that forward-backward asymmetry observable for the $B_c \to D_s l^+ l^-$ the channel is zero in the SM which consequently states parity-even nature. The non-zero value of $A_{FB}$ states parity-odd effects arising due to parity-conserving contribution coming from scalar-vector interference. $A_{FB} \neq 0$ might be possible if it receives contributions from scalar, pseudoscalar or tensor new physics operator. But in our model, no new operator has been introduced instead only the Wilson coefficients have been modified. So we stick to the zero forward-backward asymmetry and do not discuss this observable for $B_c \to D_s l^+ l^-$.

Similarly, the longitudinal polarization fraction ($P_L$) of the $D_s^*$ meson is written as [25]

$$P_L(q^2)$$
$$= \frac{H_0^{(1)}H_0^{\dagger(1)}\left(1 + \frac{4m_l^2}{q^2}\right) + H_0^{(2)}H_0^{\dagger(2)}\left(1 - \frac{4m_l^2}{q^2}\right) + \frac{2m_l^2}{q^2}3H_t^{(2)}H_t^{\dagger(2)}}{H^{(1)}H^{\dagger(1)}\left(1 + \frac{4m_l^2}{q^2}\right) + H^{(2)}H^{\dagger(2)}\left(1 - \frac{4m_l^2}{q^2}\right) + \frac{2m_l^2}{q^2}3H_t^{(2)}H_t^{\dagger(2)}}. \tag{29}$$

Here, we only investigate the longitudinal polarization of the final vector meson. The transverse polarizations $P_T$ could be obtained from the relation $P_T = 1 - P_L$. Furthermore, the leptonic polarization asymmetry ($A_{P_L}$) is defined as [63],

$$A_{P_L} = \frac{\frac{dBr_{h=-1}}{dq^2} - \frac{dBr_{h=1}}{dq^2}}{\frac{dBr_{h=-1}}{dq^2} + \frac{dBr_{h=1}}{dq^2}}$$
$$= \sqrt{1 - \frac{4m_l^2}{q^2}} \frac{2[Re\left(H_+^{(1)}H_+^{\dagger(2)}\right) + Re\left(H_-^{(1)}H_-^{\dagger(2)}\right) + Re\left(H_0^{(1)}H_0^{\dagger(2)}\right)]}{H^{(1)}H^{\dagger(1)}\left(1 + \frac{4m_l^2}{q^2}\right) + H^{(2)}H^{\dagger(2)}\left(1 - \frac{4m_l^2}{q^2}\right) + \frac{2m_l^2}{q^2}3H_t^{(2)}H_t^{\dagger(2)}}. \tag{30}$$

Here, $h$ is helicity of the final state leptons.

# V. Numerical Analysis

## A. Inputs

The input parameters which we have used for our investigation of all decay observables are as follows [64]

Table 1: Input Parameters

| | | |
|---|---|---|
| $\|V_{tb}V_{ts}\| = 0.0401 \pm 0.0010$ | $m_\mu = 0.106$ GeV | $m_b(\overline{MS}) = 4.2$ GeV |
| $M_{B_c} = 6.2751$ GeV | $m_e = 0.511 \times 10^{-3}$ GeV | $m_c(\overline{MS}) = 1.28$ GeV |
| $M_{D_s} = 1.968$ GeV | $\tau_{B_c} = 0.507 \times 10^{-12}$ s | $m_b^{pole} = 4.8$ GeV |
| $M_{D_s^*} = 2.1122$ GeV | $G_F = 1.1663787 \times 10^{-5}$ GeV$^{-2}$ | $\alpha_{em}^{-1} = 133.28$ |

The relevant Wilson coefficients have been estimated at leading order (LO) from the formulae given in sec. III and collected in Table 2.

Table 2: Wilson coefficients evaluated at renormalization scale $\mu = 4.8$ GeV

| $C_1$ | $C_2$ | $C_3$ | $C_4$ | $C_5$ | $C_6$ | $C_{7(SM)}^{eff}$ | $C_9$ | $C_{10}$ |
|---|---|---|---|---|---|---|---|---|
| -0.225 | 1.095 | 0.010 | -0.0234 | 0.007 | -0.028 | -0.304 | 4.090 | -4.484 |

A relativistic quark model based on a quasipotential approach has been used in ref. [14] to determine various form factors for the transition of $B_c \to D_s l^+ l^-$ and $B_c \to D_s^* l^+ l^-$.

It has been shown in there that the form factors have been parameterized as

$$F(q^2) = \frac{F(0)}{(1 - \frac{q^2}{M^2})(1 - \sigma_1 \frac{q^2}{M_{B_s^*}^2} + \sigma_2 \frac{q^4}{M_{B_s^*}^4})}, \quad (31)$$

for $F(q^2) = f_+(q^2), f_T(q^2), V(q^2), A_0(q^2), T_1(q^2)$ and
for $F(q^2) = f_0(q^2), A_1(q^2), A_2(q^2), T_2(q^2), T_3(q^2)$ the parameterization is as follows

$$F(q^2) = \frac{F(0)}{(1 - \sigma_1 \frac{q^2}{M_{B_s^*}^2} + \sigma_2 \frac{q^4}{M_{B_s^*}^4})}, \quad (32)$$

where $M = M_{B_s}$ for $A_0(q^2)$ and $M = M_{B_s^*}$ for all other form factors. From ref. [64] we take $M_{B_s} = 5.36689$ GeV and $M_{B_s^*} = 5.4154$ GeV to evaluate them. The form factors at $q^2 = 0$ and the fitted parameters $\sigma_1$ and $\sigma_2$ have been taken from ref. [14] and tabulated in Table 3.

Table 3: Form factors and fitted parameters for $B_c \to D_s$ and $B_c \to D_s^*$ decay mode

| | $f_+$ | $f_0$ | $f_T$ | $V$ | $A_0$ | $A_1$ | $A_2$ | $T_1$ | $T_2$ | $T_3$ |
|---|---|---|---|---|---|---|---|---|---|---|
| $F(0)$ | 0.129 | 0.129 | 0.098 | 0.182 | 0.070 | 0.089 | 0.110 | 0.085 | 0.085 | 0.051 |
| $\sigma_1$ | 2.096 | 2.331 | 1.412 | 2.133 | 1.561 | 2.479 | 2.833 | 1.540 | 2.577 | 2.783 |
| $\sigma_2$ | 1.147 | 1.666 | 0.048 | 1.183 | 0.192 | 1.686 | 2.167 | 0.248 | 1.859 | 2.170 |

These form factors provide information about the hadronization of quarks and gluons which involve QCD in the non-perturbative region and produce significant uncertainty to the decay observables. To gauge the effect of the form factor uncertainties or various observables we have taken $\pm 5\%$ error in $F(0), \sigma_1$ and $\sigma_2$ also considered the uncertainty in the CKM element.

## B. Experimental Constraints

Mainly the collider searches for production and decay of on-shell charged Higgs bosons provide the direct constraints to this new particle. Being within a much robust limit, the sensitivity of these searches is generally restricted by the extent of the allowed kinematic region of each experiment. The LEP mostly searches for the pair-production of charged Higgs bosons in Drell-Yan events, $e^+e^- \to \gamma/Z \to H^+H^-$. Considering four LEP experiments from searches in the $\tau\nu$ and $cs$ final states a lower limit of $m_{H^\pm} \gtrsim 80$ GeV has been obtained [65] assuming the absence of a light neutral Higgs boson $h^0$ in the decay $H^\pm = W^\pm h$.

At hadron colliders, searches for charged Higgs boson may be categorized into two types. First, a light charged Higgs boson below the top quark mass are being searched from top quark decays $t \to H^\pm b$; second, a charged Higgs boson could be produced in association with a top and bottom quark $pp \to H^\pm tb$. For searching a light charged Higgs boson i.e. $m_{H^+} < m_t$, the first one is more promising and has gained priority at LHC Run I with $\sqrt{s} = 7$ and 8 TeV [66-70]. For light $H^\pm$ with mass values 80 to 160 GeV the production channel via top quark decay $pp \to W^\pm H^\pm b\bar{b}$ at next-to-leading order (NLO) is considered. On the other side, LHC Run II with an increased center-of-mass energy $\sqrt{s} = 13$ TeV, $pp \to H^\pm tb$ becomes stringently important as it is sensitive to the charged Higgs boson heavier than top quark. So far LHC searches for heavy $H^\pm$ with mass range 180 GeV to 3 TeV in the production channel $pp \to H^\pm tb$ and in the charged Higgs decay modes $H^\pm \to \tau\nu$ [67, 71-73] and $H^\pm \to tb$ [74] during Run II. By CMS collaboration, the upper limits at 95% confidence level are set on $\sigma_{H^\pm} Br(H^\pm \to \tau\nu)$ for $80 < m_{H^\pm} < 3000$ GeV, including the range close to the top quark mass. The observed limit ranges from 6 pb at 80 GeV to 5 fb at 3 TeV. The results are given in ref. [75] implies the constraints in the parameter space of the MSSM benchmark scenario. In this scenario all values of $\tan\beta$ ranging from 1 to 60 remain excluded for $m_{H^\pm} \leq 160$ GeV. ATLAS collaboration [76] set 95% upper limit on $pp \to H^\pm tb$ production cross-section times branching ratio $Br(H^\pm \to tb)$ which range from 2.9 (3.0) pb at $m_{H^\pm} = 200$ GeV to 0.070 (0.077) pb at $m_{H^\pm} = 2$ TeV.

Other searches for charged Higgs boson have been performed in vector-boson fusion production channel with subsequent decay $H^\pm \to W^\pm Z$. But there is no tree level $H^\pm W^\mp Z$ coupling present in the 2HDM. In ref. [77] it is shown that while obtaining the cross-section for $pp \to H^\pm tb$ at 8 and 13 TeV, the interference contribution $\propto \lambda_{tt}\lambda_{bb}$ has been neglected and the dominating contribution in the cross-section term $\propto \lambda_{tt}^2 (\propto \lambda_{bb}^2)$ has been considered if $m_t\lambda_{tt} > m_b\lambda_{bb}$ ($m_t\lambda_{tt} < m_b\lambda_{bb}$) occurs.

The range of variations of the main free parameters i.e. $|\lambda_{tt}|$, $|\lambda_{bb}|$ and the phase angle $\theta$ in 2HDM type III are obtained from various experimental results of the electric dipole moments of neutrons, $B - \bar{B}$ mixing, $R_b = \frac{\Gamma(Z \to b\bar{b})}{\Gamma(Z \to hadrons)}$, $Br(b \to s\gamma)$ and $\rho_0 = \frac{M_W^2}{M_Z^2 \cos^2\theta}$ [78-81]. These restrictions along with the CLEO data [82] $Br(B \to X_s\gamma) = (3.15 \pm 0.35 \pm 0.32)10^{-4}$ yields $\bar{\xi}_{N,ib}^D \sim 0$, $\bar{\xi}_{N,ij}^D \sim 0$ ($i, j = d, s$) and $\bar{\xi}_{N,tc}^U \ll \bar{\xi}_{N,tt}^U$. Therefore the existence of only non-zero couplings i.e. $\bar{\xi}_{N,tt}^U$ and $\bar{\xi}_{N,bb}^D$ would be a good assumption for $b \to sl^+l^-$ transitions. LEP II obtain that for $m_{H^\pm} \geq 160$ GeV the Yukawa couplings are likely to be less than unity and $\theta$ to be in the range $60° - 90°$ with the experimental bounds on the neutron electric dipole moments and $Br(b \to s\gamma)$. On the other hand, the experimental mixing parameter $x_d = \frac{\Delta M_B}{\Gamma_B}$ where $\Delta M_B$ and $\Gamma_B$ being the mass difference and the average width for the B meson mass eigenstates constrains $|\lambda_{tt}|$ to be less than 0.3 [45]. Further limitations come from experimental results like $Br(B \to \tau\nu)$, $R_{D^{(*)}} = \frac{\Gamma(B \to D^*\mu\nu_\mu)}{\Gamma(B \to D^*l\nu_l)}$ and $Br(t \to cg)$ [77]. The experimental measurement of the parameter $R_b$ bounds the size of $|\lambda_{bb}|$ around 50 [77, 53]. This bound on $|\lambda_{bb}|$ would increase the Yukawa coupling to be greater than unity and thus hit the Landau pole at an intermediate scale ($\Lambda_{LP}$) [83]. To have finite values of all the couplings, we evaluate RGEs at the energy scale excluding $\Lambda_{LP}$. Results from CLEO and ALEPH Collaborations [84, 85] for $Br(b \to s\gamma)$ impose firm restrictions on $m_{H^\pm}$ and $\tan\beta$. Other indirect constraints on the ratio $m_{H^\pm}/\tan\beta$ could be obtained from the analysis of $B \to D\tau\bar{\nu}_\tau$ decay, where $m_{H^\pm} \geq 2.2\tan\beta$ GeV [86] and from $\tau$ lepton decays $m_{H^\pm} \geq 1.5\tan\beta$ GeV [87]. In ref. [80] one could find more restrictions on the mass of charged Higgs boson.

In ref. [37, 88] authors have studied 2HDM parameters in light of rare B decays. Branching ratio of $B_s \to \mu^+\mu^-$, $b \to s\gamma$, $(B \to K\mu^+\mu^-)^{exp}_{high\ q^2}$ and zero-crossing of the forward-backward (FB) asymmetry of $B \to K^*\mu^+\mu^-$ are used to constrain the 2HDM parameter space on $m_{H^\pm} - \tan\beta$ plane. The main reason to pick up such decay modes is that existing hadronic uncertainties in these modes are under good theoretical control. For type-I low values of $\tan\beta$ (~2) implies $m_{H^\pm} > 80$ GeV which is consistent with LEP data and also with the value of charged Higgs mass constrained from $Br(B_s \to \mu^+\mu^-)$. The zero-crossing of FB asymmetry of $B \to K^*\mu^+\mu^-$ allows the value $\tan\beta > 2.5$ but this bound decreases to 1 when $m_{H^\pm}$ is increased up to 800 GeV. In type II $m_{H^\pm} < 125$ GeV is excluded from zero-crossing of FB asymmetry. However, $Br(B \to X_s\gamma)$ narrowed down the allowed band of $m_{H^\pm}$ to $460 - 840$ GeV. Type III 2HDM has almost the same nature of constraints as that of type II except $m_{H^\pm} < 85$ GeV. Other constraints on $\lambda_{tt}$ and $\lambda_{bb}$ are extensively discussed in ref. [88]. The constraints coming from $\mathcal{B}(B_s \to \mu^+\mu^-)$ and $Br(B \to K\mu^+\mu^-)^{exp}_{high\ q^2}$ with a pull at $2.1\sigma$ level exclude the low $\tan\beta \lesssim 1$ region irrespective of the type of 2HDM. Further discussion of this bound could be found in ref. [77].

Combining electroweak precision data, Higgs coupling measurements, flavor observables and the anomalous muonic magnetic moment, global fit has been done on parameters of different types of 2HDM [42]. The lightest scalar Higgs boson $m_{h^0}$ is identified with the observed SM Higgs boson of mass $125.09 \pm 0.24$ GeV. While no other NP contribution except 2HDM is assumed, all other model parameters are allowed to vary as $130 < m_{H^0}, m_{A^0} < 1000$ GeV, $100 < m_{H^\pm} < 1000$ GeV, $0 \leq \beta - \alpha \leq \pi$, $0.001 < \tan\beta < 50$ and $-8 \times 10^5 < m_{12}^2 < 8 \times 10^5$. Due to huge range of 2HDM parameters, these fitting values provide only weak exclusion limits on scalar masses. From the searches of LEP [65] experiments a lower limit is reported for type I scenario as $m_{H^\pm} > 72.5$ GeV, whereas LHC searches [77] have found the limit for type II scenario as $m_{H^\pm} \gtrsim 150$ GeV. Stronger bounds on mass limits mainly on $m_{H^\pm}$ could be attained for a specific region of $\tan\beta$.

### C. Decay observables for $B_c \to (D_s, D_s^*)l^+l^-$ in 2HDM

In this paper, we have studied the branching ratio, lepton polarization asymmetry, forward-backward asymmetry, longitudinal polarization fraction and LFU ratios for $B_c \to (D_s, D_s^*)\mu^+\mu^-$ channels. We first calculate these parameters in the SM, results are shown in

Table 4 – Table 7. The observables have been evaluated for three different $q^2$ bins i.e. lower region (0.045-1), middle region (1-6) and higher region (13.6-17.5). We have excluded the region for $c\bar{c}$ loop resonance in our numerical analysis.

Table 4: Branching ratio of $B_c \to (D_s, D_s^*)\mu^+\mu^-$ ($Br \times 10^{-8}$) in the SM.

| Decay mode | (0.045-1) GeV$^2$ | (1-6) GeV$^2$ | (13.6-17.5) GeV$^2$ |
|---|---|---|---|
| $B_c \to D_s\mu^+\mu^-$ | $0.274 \pm 0.029$ | $2.204 \pm 0.266$ | $6.504 \pm 2.986$ |
| $B_c \to D_s^*\mu^+\mu^-$ | $0.226 \pm 0.028$ | $1.273 \pm 0.255$ | $18.859 \pm 12.955$ |

Table 5: Lepton polarization asymmetry ($A_{P_L}$) of $B_c \to (D_s, D_s^*)\mu^+\mu^-$ in the SM.

| Decay mode | (0.045-1) GeV$^2$ | (1-6) GeV$^2$ | (13.6-17.5) GeV$^2$ |
|---|---|---|---|
| $B_c \to D_s\mu^+\mu^-$ | $-0.783 \pm 0.007$ | $-0.967 \pm 0.001$ | $-0.762 \pm 0.007$ |
| $B_c \to D_s^*\mu^+\mu^-$ | $-0.159 \pm 0.090$ | $-0.864 \pm 0.031$ | $-0.674 \pm 0.049$ |

Table 6: Lepton flavor universality (LFU) ratio: $R_{D_s}$ for $B_c \to D_s\mu^+\mu^-$ and $R_{D_s^*}$ for $B_c \to D_s^*\mu^+\mu^-$ in the SM.

| Decay mode | (1-6) GeV$^2$ | (13.6-17.5) GeV$^2$ |
|---|---|---|
| $B_c \to D_s\mu^+\mu^-$ | $1.005 \pm 0.001$ | $0.976 \pm 0.003$ |
| $B_c \to D_s^*\mu^+\mu^-$ | $0.996 \pm 0.001$ | $0.973 \pm 0.000$ |

Table 7: Forward-backward asymmetry ($A_{FB}$) and polarization fraction ($P_L$) of $B_c \to D_s^*\mu^+\mu^-$ in the SM.

| Decay mode | Parameter | (0.045-1) GeV$^2$ | (1-6) GeV$^2$ | (13.6-17.5) GeV$^2$ |
|---|---|---|---|---|
| $B_c \to D_s^*\mu^+\mu^-$ | $A_{FB}$ | $-0.080 \pm 0.011$ | $0.107 \pm 0.019$ | $0.130 \pm 0.071$ |
| | $P_L$ | $0.417 \pm 0.063$ | $0.591 \pm 0.057$ | $0.293 \pm 0.081$ |

The central values of observables are determined from the central values of input parameters and propagating the error through form factors and CKM elements, we have got the uncertainties in decay observables. Within the SM, the branching ratio for both $B_c \to D_s$ and $B_c \to D_s^*$ transitions are found to be $\sim 10^{-8}$ whereas we obtain one order increment in high $q^2$ region for the channel $B_c \to D_s^*\mu^+\mu^-$. These results are expected to be within the sensitivity of future run of LHCb because of the plentiful production of $B_c$ meson. As discussed before for the pseudo-scalar $D$ mesons, there is no forward-backward asymmetry present in our analysis due to zero transverse helicity. So $D$ mesons do not have any polarization direction resulting unit value in the longitudinal polarization fraction. In the SM, the LFU parameters $R_{D_s}, R_{D_s^*} \sim 1$ within the region of $1 < q^2 < 6$ GeV$^2$. From Tables 4-7, it is very clear that the uncertainties associated with LFU ratios are very less than other observables.

Our main interest is to determine the effect of charged Higgs boson on decay observables of $B_c \to (D_s, D_s^*)\mu^+\mu^-$ modes in type III 2HDM. To this end, we have calculated all these observables in type III 2HDM considering $\theta = \frac{\pi}{2}$ and $\lambda_{bb} = 50$. All the measurements are done for different mass of charged Higgs boson within the region $160 \text{ GeV} < m_{H^\pm} < 1000 \text{ GeV}$. All the results in 2HDM (type III) are tabulated in Tables 8-15.

Table 8 (a): Branching ratio of $B_c \to D_s \mu^+ \mu^-$ ($Br \times 10^{-8}$) in 2HDM (type III) with $\lambda_{tt} = 0.05$.

| $q^2$ GeV$^2$ | $m_{H^\pm} = 160$ GeV | $m_{H^\pm} = 250$ GeV | $m_{H^\pm} = 500$ GeV | $m_{H^\pm} = 1000$ GeV |
|---|---|---|---|---|
| 0.045-1 | $0.268 \pm 0.029$ | $0.267 \pm 0.029$ | $0.267 \pm 0.029$ | $0.266 \pm 0.029$ |
| 1-6 | $2.162 \pm 0.264$ | $2.158 \pm 0.263$ | $2.154 \pm 0.263$ | $2.151 \pm 0.263$ |
| 13.6-17.5 | $6.463 \pm 2.976$ | $6.463 \pm 2.975$ | $6.464 \pm 2.974$ | $6.465 \pm 2.973$ |

Table 8 (b): Branching ratio of $B_c \to D_s \mu^+ \mu^-$ ($Br \times 10^{-8}$) in 2HDM (type III) with $\lambda_{tt} = 0.15$.

| $q^2$ GeV$^2$ | $m_{H^\pm} = 160$ GeV | $m_{H^\pm} = 250$ GeV | $m_{H^\pm} = 500$ GeV | $m_{H^\pm} = 1000$ GeV |
|---|---|---|---|---|
| 0.045-1 | $0.282 \pm 0.030$ | $0.277 \pm 0.030$ | $0.271 \pm 0.029$ | $0.268 \pm 0.029$ |
| 1-6 | $2.272 \pm 0.275$ | $2.232 \pm 0.271$ | $2.188 \pm 0.267$ | $2.165 \pm 0.264$ |
| 13.6-17.5 | $6.583 \pm 3.042$ | $6.543 \pm 3.023$ | $6.501 \pm 2.998$ | $6.480 \pm 2.983$ |

Table 8 (c): Branching ratio of $B_c \to D_s \mu^+ \mu^-$ ($Br \times 10^{-8}$) in 2HDM (type III) with $\lambda_{tt} = 0.3$.

| $q^2$ GeV$^2$ | $m_{H^\pm} = 160$ GeV | $m_{H^\pm} = 250$ GeV | $m_{H^\pm} = 500$ GeV | $m_H = 1000$ GeV |
|---|---|---|---|---|
| 0.045-1 | $0.334 \pm 0.035$ | $0.312 \pm 0.033$ | $0.287 \pm 0.031$ | $0.274 \pm 0.029$ |
| 1-6 | $2.658 \pm 0.315$ | $2.492 \pm 0.300$ | $2.309 \pm 0.281$ | $2.213 \pm 0.270$ |
| 13.6-17.5 | $7.107 \pm 3.307$ | $6.890 \pm 3.210$ | $6.659 \pm 3.090$ | $6.542 \pm 3.021$ |

Table 9 (a): Lepton polarization asymmetry ($A_{P_L}$) of $B_c \to D_s \mu^+ \mu^-$ in 2HDM (type III) with $\lambda_{tt} = 0.05$.

| $q^2$ GeV$^2$ | $m_{H^\pm} = 160$ GeV | $m_{H^\pm} = 250$ GeV | $m_{H^\pm} = 500$ GeV | $m_{H^\pm} = 1000$ GeV |
|---|---|---|---|---|
| 0.045-1 | $-0.784 \pm 0.008$ | $-0.785 \pm 0.008$ | $-0.785 \pm 0.008$ | $-0.785 \pm 0.008$ |
| 1-6 | $-0.970 \pm 0.001$ | $-0.970 \pm 0.001$ | $-0.970 \pm 0.001$ | $-0.970 \pm 0.001$ |
| 13.6-17.5 | $-0.758 \pm 0.009$ | $-0.756 \pm 0.009$ | $-0.755 \pm 0.010$ | $-0.753 \pm 0.010$ |

Table 9 (b): Lepton polarization asymmetry ($A_{P_L}$) of $B_c \to D_s \mu^+ \mu^-$ in 2HDM (type III) with $\lambda_{tt} = 0.15$.

| $q^2$ GeV$^2$ | $m_{H^\pm} = 160$ GeV | $m_{H^\pm} = 250$ GeV | $m_{H^\pm} = 500$ GeV | $m_{H^\pm} = 1000$ GeV |
|---|---|---|---|---|
| 0.045-1 | $-0.777 \pm 0.008$ | $-0.782 \pm 0.008$ | $-0.785 \pm 0.008$ | $-0.785 \pm 0.008$ |
| 1-6 | $-0.963 \pm 0.002$ | $-0.967 \pm 0.002$ | $-0.970 \pm 0.002$ | $-0.970 \pm 0.001$ |
| 13.6-17.5 | $-0.765 \pm 0.010$ | $-0.763 \pm 0.009$ | $-0.759 \pm 0.009$ | $-0.755 \pm 0.010$ |

Table 9 (c): Lepton polarization asymmetry ($A_{P_L}$) of $B_c \to D_s \mu^+ \mu^-$ in 2HDM (type III) with $\lambda_{tt} = 0.3$.

| $q^2$ GeV$^2$ | $m_{H^\pm} = 160$ GeV | $m_{H^\pm} = 250$ GeV | $m_{H^\pm} = 500$ GeV | $m_{H^\pm} = 1000$ GeV |
|---|---|---|---|---|
| 0.045-1 | $-0.757 \pm 0.010$ | $-0.771 \pm 0.009$ | $-0.782 \pm 0.008$ | $-0.785 \pm 0.008$ |
| 1-6 | $-0.944 \pm 0.005$ | $-0.958 \pm 0.003$ | $-0.968 \pm 0.002$ | $-0.970 \pm 0.001$ |
| 13.6-17.5 | $-0.764 \pm 0.019$ | $-0.771 \pm 0.013$ | $-0.766 \pm 0.009$ | $-0.759 \pm 0.009$ |

Table 10 (a): Lepton non-universality parameter ($R_{D_s}$) of $B_c \to D_s \mu^+ \mu^-$ in 2HDM (type III) with $\lambda_{tt} = 0.05$.

| $q^2$ GeV² | $m_{H^\pm} = 160$ GeV | $m_{H^\pm} = 250$ GeV | $m_{H^\pm} = 500$ GeV | $m_{H^\pm} = 1000$ GeV |
|---|---|---|---|---|
| 1-6 | 0.986 ± 0.002 | 0.984 ± 0.002 | 0.982 ± 0.002 | 0.981 ± 0.002 |
| 13.6-17.5 | 0.960 ± 0.006 | 0.960 ± 0.006 | 0.960 ± 0.006 | 0.960 ± 0.006 |

Table 10 (b): Lepton non-universality parameter ($R_{D_s}$) of $B_c \to D_s \mu^+ \mu^-$ in 2HDM (type III) with $\lambda_{tt} = 0.15$.

| $q^2$ GeV² | $m_{H^\pm} = 160$ GeV | $m_{H^\pm} = 250$ GeV | $m_{H^\pm} = 500$ GeV | $m_{H^\pm} = 1000$ GeV |
|---|---|---|---|---|
| 1-6 | 1.036 ± 0.002 | 1.018 ± 0.002 | 0.998 ± 0.002 | 0.987 ± 0.002 |
| 13.6-17.5 | 1.003 ± 0.005 | 0.989 ± 0.006 | 0.973 ± 0.006 | 0.965 ± 0.006 |

Table 10 (c): Lepton non-universality parameter ($R_{D_s}$) of $B_c \to D_s \mu^+ \mu^-$ in 2HDM (type III) with $\lambda_{tt} = 0.3$.

| $q^2$ GeV² | $m_{H^\pm} = 160$ GeV | $m_{H^\pm} = 250$ GeV | $m_{H^\pm} = 500$ GeV | $m_{H^\pm} = 1000$ GeV |
|---|---|---|---|---|
| 1-6 | 1.214 ± 0.005 | 1.137 ± 0.003 | 1.053 ± 0.002 | 1.009 ± 0.002 |
| 13.6-17.5 | 1.183 ± 0.012 | 1.109 ± 0.003 | 1.028 ± 0.006 | 0.987 ± 0.006 |

Table 11 (a): Branching ratio of $B_c \to D_s^* \mu^+ \mu^-$ ($Br \times 10^{-8}$) in 2HDM (type III) with $\lambda_{tt} = 0.05$.

| $q^2$ GeV² | $m_{H^\pm} = 160$ GeV | $m_{H^\pm} = 250$ GeV | $m_{H^\pm} = 500$ GeV | $m_{H^\pm} = 1000$ GeV |
|---|---|---|---|---|
| 0.045-1 | 0.774 ± 0.069 | 0.625 ± 0.057 | 0.517 ± 0.049 | 0.489 ± 0.046 |
| 1-6 | 1.397 ± 0.239 | 1.277 ± 0.236 | 1.189 ± 0.235 | 1.166 ± 0.234 |
| 13.6-17.5 | 18.518 ± 12.650 | 18.500 ± 12.673 | 18.517 ± 12.698 | 18.544 ± 12.710 |

Table 11 (b): Branching ratio of $B_c \to D_s^* \mu^+ \mu^-$ ($Br \times 10^{-8}$) in 2HDM (type III) with $\lambda_{tt} = 0.15$.

| $q^2$ GeV² | $m_{H^\pm} = 160$ GeV | $m_{H^\pm} = 250$ GeV | $m_{H^\pm} = 500$ GeV | $m_{H^\pm} = 1000$ GeV |
|---|---|---|---|---|
| 0.045-1 | 3.092 ± 0.270 | 1.753 ± 0.153 | 0.777 ± 0.070 | 0.532 ± 0.050 |
| 1-6 | 3.319 ± 0.344 | 2.222 ± 0.277 | 1.415 ± 0.242 | 1.207 ± 0.236 |
| 13.6-17.5 | 19.713 ± 12.797 | 19.061 ± 12.769 | 18.632 ± 12.754 | 18.557 ± 12.737 |

Table 11 (c): Branching ratio of $B_c \to D_s^* \mu^+ \mu^-$ ($Br \times 10^{-8}$) in 2HDM (type III) with $\lambda_{tt} = 0.3$.

| $q^2$ GeV² | $m_{H^\pm} = 160$ GeV | $m_{H^\pm} = 250$ GeV | $m_{H^\pm} = 500$ GeV | $m_{H^\pm} = 1000$ GeV |
|---|---|---|---|---|
| 0.045-1 | 10.092 ± 0.957 | 5.564 ± 0.487 | 1.658 ± 0.145 | 0.676 ± 0.062 |
| 1-6 | 9.833 ± 0.890 | 5.432 ± 0.510 | 2.190 ± 0.281 | 1.349 ± 0.244 |
| 13.6-17.5 | 24.846 ± 14.516 | 21.706 ± 13.592 | 19.368 ± 13.092 | 18.738 ± 12.881 |

Table 12 (a): Forward-backward asymmetry ($A_{FB}$) of $B_c \to D_s^* \mu^+ \mu^-$ in 2HDM (type III) with $\lambda_{tt} = 0.05$.

| $q^2$ GeV² | $m_{H^\pm} = 160$ GeV | $m_{H^\pm} = 250$ GeV | $m_{H^\pm} = 500$ GeV | $m_{H^\pm} = 1000$ GeV |
|---|---|---|---|---|
| 0.045-1 | −0.048 ± 0.003 | −0.060 ± 0.004 | −0.073 ± 0.018 | −0.077 ± 0.006 |
| 1-6 | 0.048 ± 0.013 | 0.047 ± 0.016 | 0.045 ± 0.018 | 0.045 ± 0.019 |
| 13.6-17.5 | 0.124 ± 0.069 | 0.124 ± 0.070 | 0.123 ± 0.069 | 0.123 ± 0.069 |

Table 12 (b): Forward-backward asymmetry ($A_{FB}$) of $B_c \to D_s^* \mu^+ \mu^-$ in 2HDM (type III) with $\lambda_{tt} = 0.15$.

| $q^2$ GeV$^2$ | $m_{H^\pm} = 160$ GeV | $m_{H^\pm} = 250$ GeV | $m_{H^\pm} = 500$ GeV | $m_{H^\pm} = 1000$ GeV |
|---|---|---|---|---|
| 0.045-1 | $-0.012 \pm 0.001$ | $-0.021 \pm 0.001$ | $-0.048 \pm 0.003$ | $-0.071 \pm 0.005$ |
| 1-6 | $0.035 \pm 0.006$ | $0.044 \pm 0.008$ | $0.049 \pm 0.013$ | $0.046 \pm 0.018$ |
| 13.6-17.5 | $0.115 \pm 0.058$ | $0.121 \pm 0.064$ | $0.124 \pm 0.069$ | $0.123 \pm 0.070$ |

Table 12 (c): Forward-backward asymmetry ($A_{FB}$) of $B_c \to D_s^* \mu^+ \mu^-$ in 2HDM (type III) with $\lambda_{tt} = 0.3$.

| $q^2$ GeV$^2$ | $m_{H^\pm} = 160$ GeV | $m_{H^\pm} = 250$ GeV | $m_{H^\pm} = 500$ GeV | $m_{H^\pm} = 1000$ GeV |
|---|---|---|---|---|
| 0.045-1 | $-0.004 \pm 0.000$ | $-0.007 \pm 0.000$ | $-0.023 \pm 0.001$ | $-0.056 \pm 0.004$ |
| 1-6 | $0.018 \pm 0.003$ | $0.028 \pm 0.005$ | $0.046 \pm 0.008$ | $0.049 \pm 0.015$ |
| 13.6-17.5 | $0.091 \pm 0.048$ | $0.107 \pm 0.052$ | $0.123 \pm 0.065$ | $0.125 \pm 0.069$ |

Table 13 (a): Polarization fraction ($P_L$) of $B_c \to D_s^* \mu^+ \mu^-$ in 2HDM (type III) with $\lambda_{tt} = 0.05$.

| $q^2$ GeV$^2$ | $m_{H^\pm} = 160$ GeV | $m_{H^\pm} = 250$ GeV | $m_{H^\pm} = 500$ GeV | $m_{H^\pm} = 1000$ GeV |
|---|---|---|---|---|
| 0.045-1 | $0.131 \pm 0.035$ | $0.162 \pm 0.041$ | $0.195 \pm 0.048$ | $0.206 \pm 0.050$ |
| 1-6 | $0.458 \pm 0.070$ | $0.505 \pm 0.070$ | $0.549 \pm 0.069$ | $0.561 \pm 0.069$ |
| 13.6-17.5 | $0.293 \pm 0.083$ | $0.293 \pm 0.083$ | $0.294 \pm 0.083$ | $0.294 \pm 0.083$ |

Table 13 (b): Polarization fraction ($P_L$) of $B_c \to D_s^* \mu^+ \mu^-$ in 2HDM (type III) with $\lambda_{tt} = 0.15$.

| $q^2$ GeV$^2$ | $m_{H^\pm} = 160$ GeV | $m_{H^\pm} = 250$ GeV | $m_{H^\pm} = 500$ GeV | $m_{H^\pm} = 1000$ GeV |
|---|---|---|---|---|
| 0.045-1 | $0.037 \pm 0.010$ | $0.062 \pm 0.017$ | $0.132 \pm 0.035$ | $0.191 \pm 0.047$ |
| 1-6 | $0.223 \pm 0.046$ | $0.305 \pm 0.058$ | $0.458 \pm 0.070$ | $0.542 \pm 0.069$ |
| 13.6-17.5 | $0.286 \pm 0.079$ | $0.290 \pm 0.082$ | $0.293 \pm 0.083$ | $0.294 \pm 0.083$ |

Table 13 (c): Polarization fraction ($P_L$) of $B_c \to D_s^* \mu^+ \mu^-$ in 2HDM (type III) with $\lambda_{tt} = 0.3$.

| $q^2$ GeV$^2$ | $m_{H^\pm} = 160$ GeV | $m_{H^\pm} = 250$ GeV | $m_{H^\pm} = 500$ GeV | $m_{H^\pm} = 1000$ GeV |
|---|---|---|---|---|
| 0.045-1 | $0.015 \pm 0.003$ | $0.025 \pm 0.006$ | $0.067 \pm 0.019$ | $0.154 \pm 0.040$ |
| 1-6 | $0.114 \pm 0.022$ | $0.165 \pm 0.034$ | $0.319 \pm 0.060$ | $0.490 \pm 0.070$ |
| 13.6-17.5 | $0.272 \pm 0.065$ | $0.281 \pm 0.074$ | $0.290 \pm 0.082$ | $0.293 \pm 0.083$ |

Table 14 (a): Lepton polarization asymmetry ($A_{P_L}$) of $B_c \to D_s^* \mu^+ \mu^-$ in 2HDM (type III) with $\lambda_{tt} = 0.05$.

| $q^2$ GeV$^2$ | $m_{H^\pm} = 160$ GeV | $m_{H^\pm} = 250$ GeV | $m_{H^\pm} = 500$ GeV | $m_{H^\pm} = 1000$ GeV |
|---|---|---|---|---|
| 0.045-1 | $0.012 \pm 0.037$ | $0.015 \pm 0.046$ | $0.019 \pm 0.056$ | $0.020 \pm 0.059$ |
| 1-6 | $-0.569 \pm 0.068$ | $-0.614 \pm 0.069$ | $-0.654 \pm 0.069$ | $-0.665 \pm 0.069$ |
| 13.6-17.5 | $-0.622 \pm 0.097$ | $-0.622 \pm 0.093$ | $-0.620 \pm 0.091$ | $-0.617 \pm 0.092$ |

Table 14 (b): Lepton polarization asymmetry ($A_{P_L}$) of $B_c \to D_s^* \mu^+ \mu^-$ in 2HDM (type III) with $\lambda_{tt} = 0.15$.

| $q^2$ GeV$^2$ | $m_{H^\pm} = 160$ GeV | $m_{H^\pm} = 250$ GeV | $m_{H^\pm} = 500$ GeV | $m_{H^\pm} = 1000$ GeV |
|---|---|---|---|---|
| 0.045-1 | $0.002 \pm 0.001$ | $0.005 \pm 0.016$ | $0.011 \pm 0.037$ | $0.018 \pm 0.054$ |
| 1-6 | $-0.291 \pm 0.052$ | $-0.400 \pm 0.061$ | $-0.571 \pm 0.068$ | $-0.649 \pm 0.069$ |
| 13.6-17.5 | $-0.579 \pm 0.170$ | $-0.609 \pm 0.128$ | $-0.624 \pm 0.096$ | $-0.621 \pm 0.091$ |

Table 14 (c): Lepton polarization asymmetry ($A_{P_L}$) of $B_c \to D_s^* \mu^+ \mu^-$ in 2HDM (type III) with $\lambda_{tt} = 0.3$.

| $q^2$ GeV$^2$ | $m_{H^\pm} = 160$ GeV | $m_{H^\pm} = 250$ GeV | $m_{H^\pm} = 500$ GeV | $m_{H^\pm} = 1000$ GeV |
|---|---|---|---|---|
| 0.045-1 | $0.000 \pm 0.003$ | $0.004 \pm 0.006$ | $0.004 \pm 0.018$ | $0.012 \pm 0.043$ |
| 1-6 | $-0.131 \pm 0.029$ | $-0.211 \pm 0.042$ | $-0.420 \pm 0.062$ | $-0.604 \pm 0.068$ |
| 13.6-17.5 | $-0.464 \pm 0.274$ | $-0.544 \pm 0.215$ | $-0.618 \pm 0.121$ | $-0.627 \pm 0.092$ |

Table 15 (a): Lepton non-universality parameter ($R_{D_s^*}$) of $B_c \to D_s^* \mu^+ \mu^-$ in 2HDM (type III) with $\lambda_{tt} = 0.05$.

| $q^2$ GeV$^2$ | $m_{H^\pm} = 160$ GeV | $m_{H^\pm} = 250$ GeV | $m_{H^\pm} = 500$ GeV | $m_{H^\pm} = 1000$ GeV |
|---|---|---|---|---|
| 1-6 | $1.054 \pm 0.015$ | $1.035 \pm 0.012$ | $1.018 \pm 0.010$ | $1.013 \pm 0.009$ |
| 13.6-17.5 | $0.972 \pm 0.004$ | $0.971 \pm 0.003$ | $0.971 \pm 0.003$ | $0.971 \pm 0.002$ |

Table 15 (b): Lepton non-universality parameter ($R_{D_s^*}$) of $B_c \to D_s^* \mu^+ \mu^-$ in 2HDM (type III) with $\lambda_{tt} = 0.15$.

| $q^2$ GeV$^2$ | $m_{H^\pm} = 160$ GeV | $m_{H^\pm} = 250$ GeV | $m_{H^\pm} = 500$ GeV | $m_{H^\pm} = 1000$ GeV |
|---|---|---|---|---|
| 1-6 | $1.163 \pm 0.019$ | $1.122 \pm 0.019$ | $1.053 \pm 0.014$ | $1.020 \pm 0.010$ |
| 13.6-17.5 | $0.990 \pm 0.022$ | $0.980 \pm 0.012$ | $0.972 \pm 0.004$ | $0.971 \pm 0.003$ |

Table 15 (c): Lepton non-universality parameter ($R_{D_s^*}$) of $B_c \to D_s^* \mu^+ \mu^-$ in 2HDM (type III) with $\lambda_{tt} = 0.3$.

| $q^2$ GeV$^2$ | $m_{H^\pm} = 160$ GeV | $m_{H^\pm} = 250$ GeV | $m_{H^\pm} = 500$ GeV | $m_{H^\pm} = 1000$ GeV |
|---|---|---|---|---|
| 1-6 | $1.225 \pm 0.015$ | $1.195 \pm 0.017$ | $1.115 \pm 0.018$ | $1.040 \pm 0.013$ |
| 13.6-17.5 | $1.026 \pm 0.043$ | $1.003 \pm 0.031$ | $0.979 \pm 0.011$ | $0.971 \pm 0.004$ |

The graphical explanations of differential branching ratio and other decay observables (for both $B_c \to D_s \mu^+ \mu^-$ and $B_c \to D_s^* \mu^+ \mu^-$ channels) with model parameters in type I, II and III 2HDM are shown in figures below. In order to show the dependence of decay observables on $\tan \beta$ as well as $\lambda_{tt}$ we have depicted the decay distribution plots in all three types of 2HDM. As the mass spectrum and experimental bounds are different for these three scenarios, so numerical values would also be different. But concerning the length of this paper final values are given only for the type III model.

In the figures, the solid black line represents the SM prediction whereas the dashed lines are labeled for different mass of charged Higgs boson as: black solid horizontal line- SM prediction, blue dashed- $m_{H^\pm} = 160$ GeV, red dashed- $m_{H^\pm} = 250$ GeV, green dashed- $m_{H^\pm} = 500$ GeV, purple dashed- $m_{H^\pm} = 1000$ GeV.

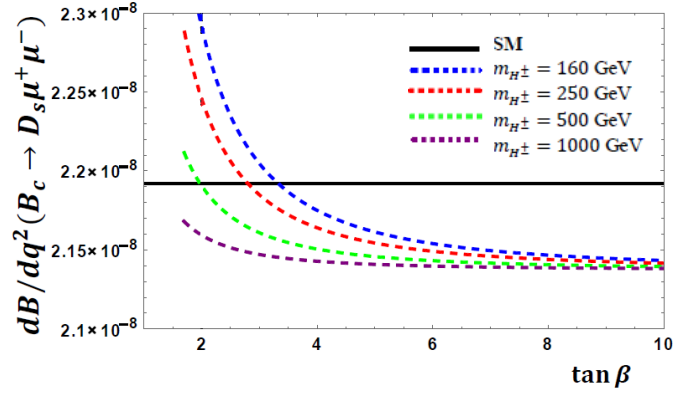

Fig. 3 (a): Variation of differential branching ratio of $B_c \to D_s \mu^+ \mu^-$ with $\tan\beta$ in type I 2HDM

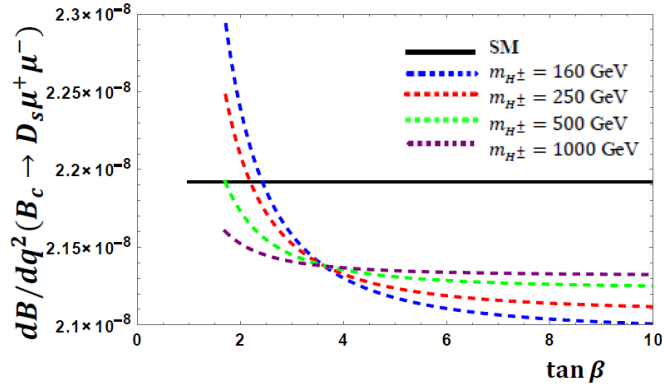

Fig. 3 (b): Variation of differential branching ratio of $B_c \to D_s \mu^+ \mu^-$ with $\tan\beta$ in type II 2HDM

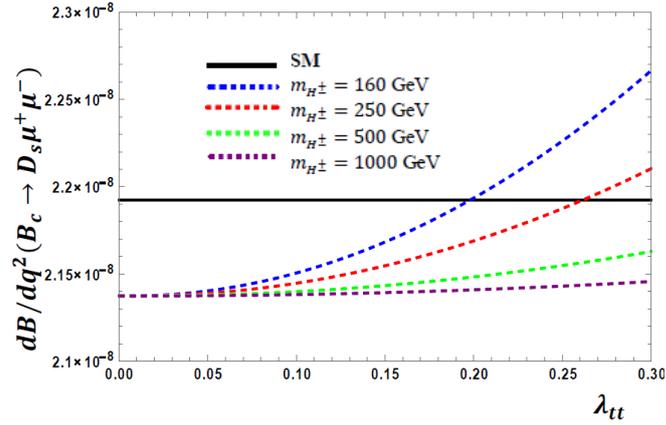

Fig. 3 (c): Variation of differential branching ratio of $B_c \to D_s \mu^+ \mu^-$ with $\lambda_{tt}$ in type III 2HDM

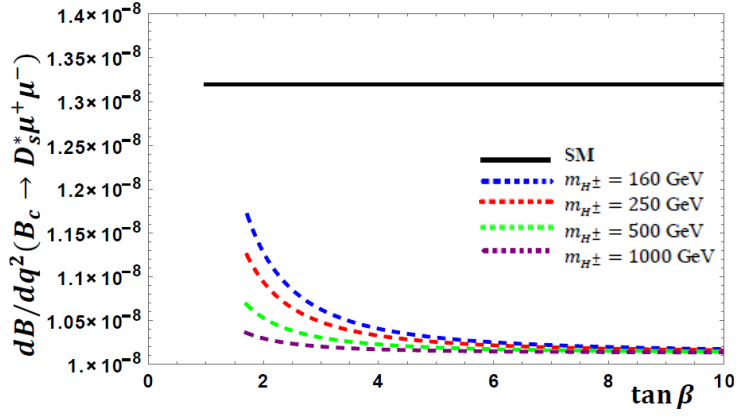

Fig. 4 (a): Variation of differential branching ratio of $B_c \to D_s^* \mu^+ \mu^-$ with $\tan\beta$ in type I 2HDM

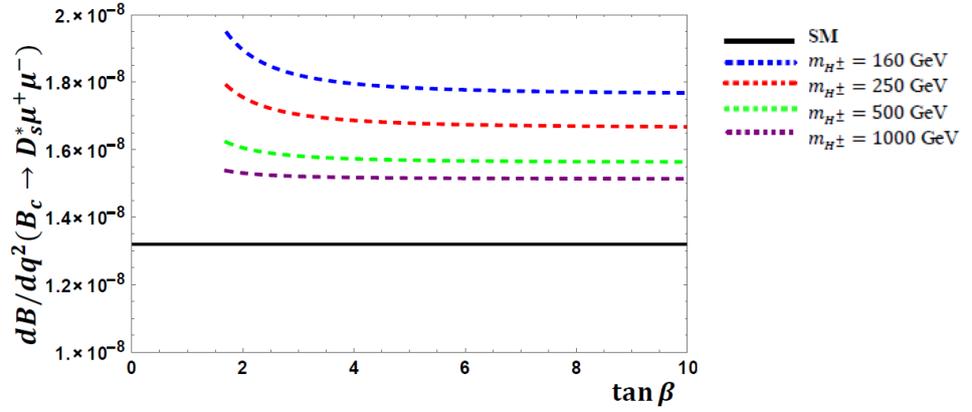

Fig. 4 (b): Variation of differential branching ratio of $B_c \to D_s^* \mu^+ \mu^-$ with $\tan\beta$ in type II 2HDM

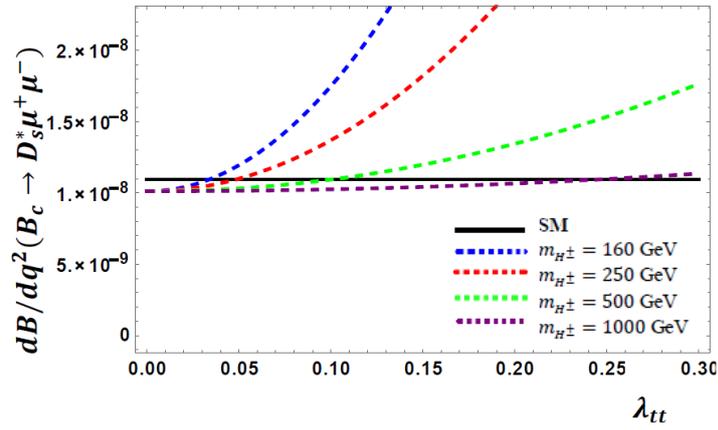

Fig. 4 (c): Variation of differential branching ratio of $B_c \to D_s^* \mu^+ \mu^-$ with $\lambda_{tt}$ in type III 2HDM

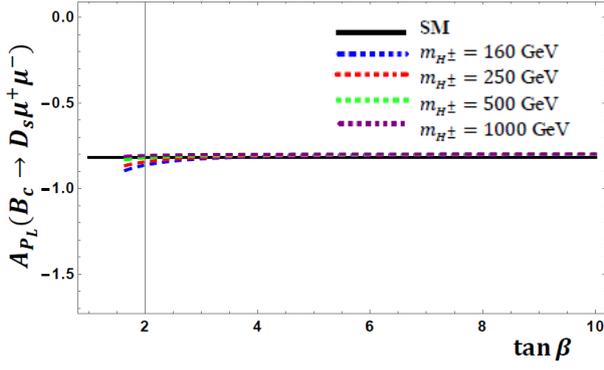

Fig. 5 (a): Variation of lepton polarization asymmetry ($A_{P_L}$) of $B_c \to D_s \mu^+ \mu^-$ with $\tan\beta$ in type I 2HDM

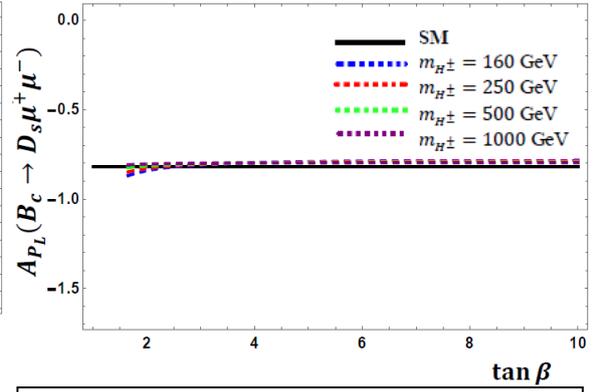

Fig. 5 (b): Variation of lepton polarization asymmetry ($A_{P_L}$) of $B_c \to D_s \mu^+ \mu^-$ with $\tan\beta$ in type II 2HDM

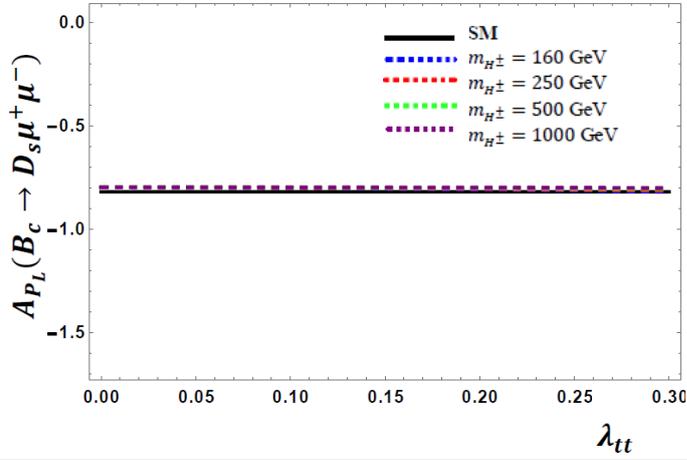

Fig. 5 (c): Variation of lepton polarization asymmetry ($A_{P_L}$) of $B_c \to D_s \mu^+ \mu^-$ with $\lambda_{tt}$ in type III 2HDM

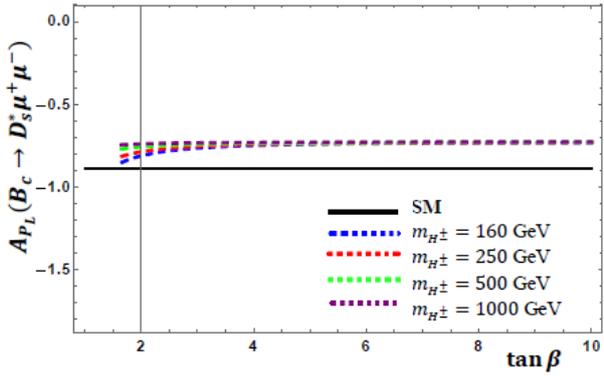

Fig. 6 (a): Variation of lepton polarization asymmetry ($A_{P_L}$) of $B_c \to D_s^* \mu^+ \mu^-$ with $\tan\beta$ in type I 2HDM

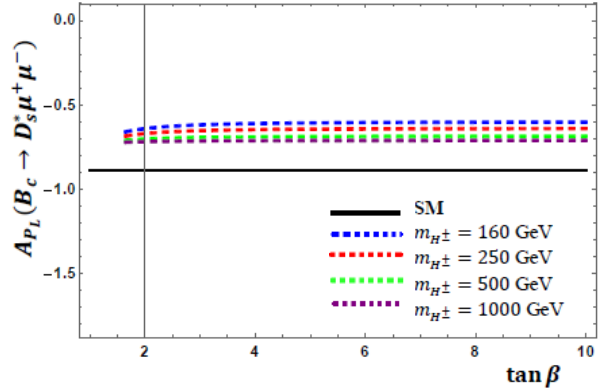

Fig. 6 (b): Variation of lepton polarization asymmetry ($A_{P_L}$) of $B_c \to D_s^* \mu^+ \mu^-$ with $\tan\beta$ in type II 2HDM

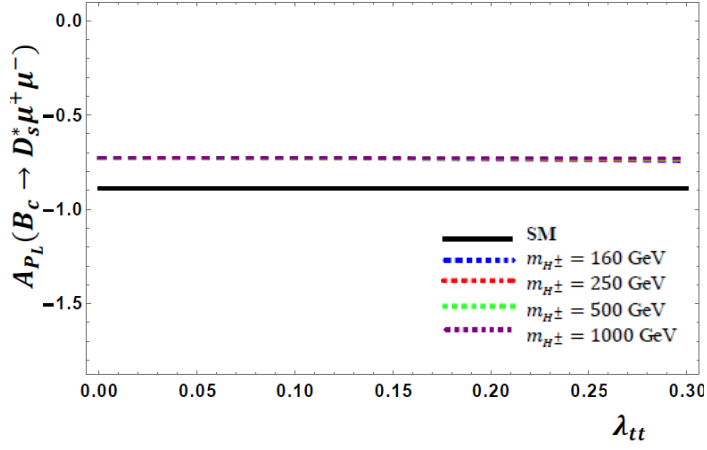

Fig. 6 (c): Variation of lepton polarization asymmetry ($A_{P_L}$) of $B_c \to D_s^* \mu^+ \mu^-$ with $\lambda_{tt}$ in type III 2HDM

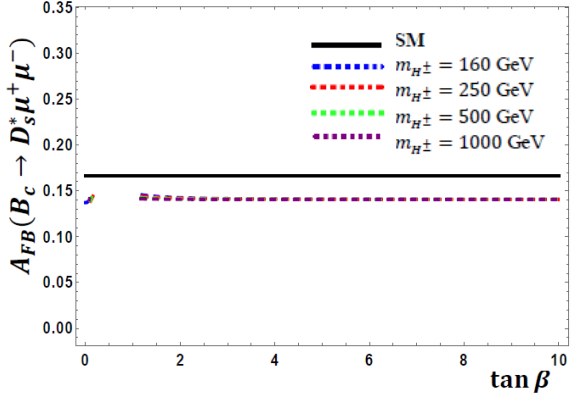

Fig. 7 (a): Variation of forward-backward asymmetry ($A_{FB}$) of $B_c \to D_s^* \mu^+ \mu^-$ with $\tan\beta$ in type I 2HDM

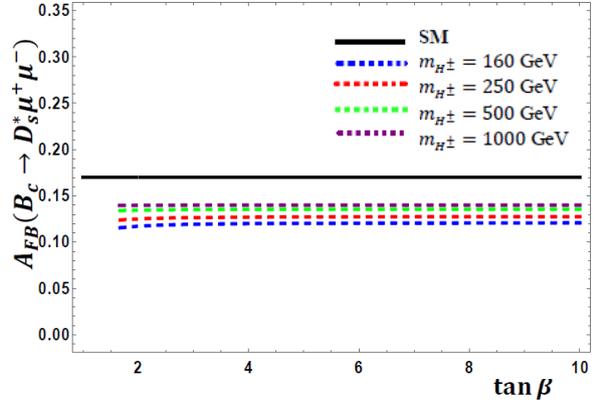

Fig. 7 (b): Variation of forward-backward asymmetry ($A_{FB}$) of $B_c \to D_s^* \mu^+ \mu^-$ with $\tan\beta$ in type II 2HDM

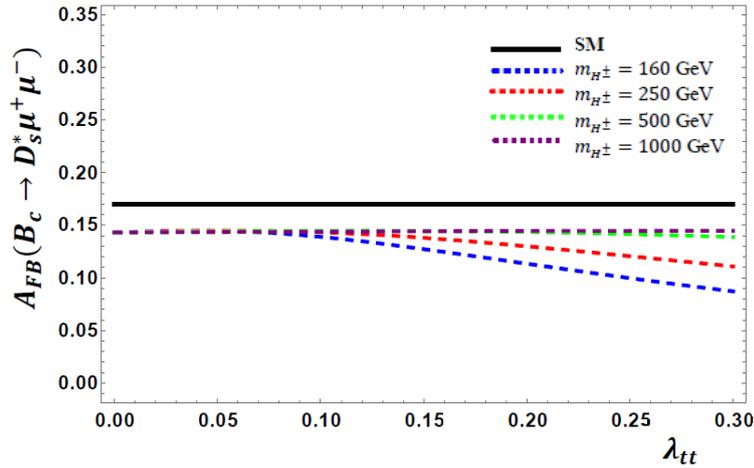

Fig. 7 (c): Variation of forward-backward asymmetry ($A_{FB}$) of $B_c \to D_s^* \mu^+ \mu^-$ with $\lambda_{tt}$ in type III 2HDM

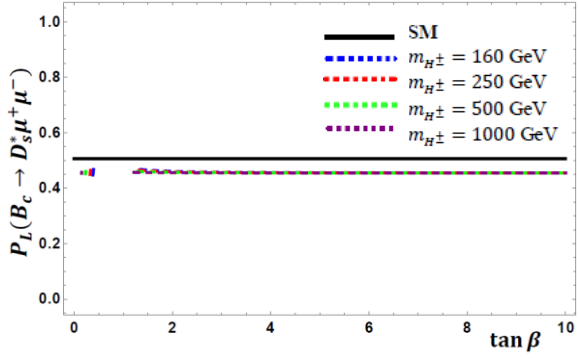

Fig. 8 (a): Variation of polarization fraction ($P_L$) of $B_c \to D_s^* \mu^+ \mu^-$ with $\tan\beta$ in type I 2HDM

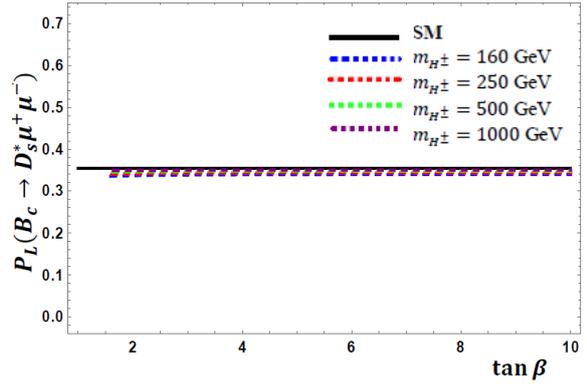

Fig. 8 (b): Variation of polarization fraction ($P_L$) of $B_c \to D_s^* \mu^+ \mu^-$ with $\tan\beta$ in type II 2HDM

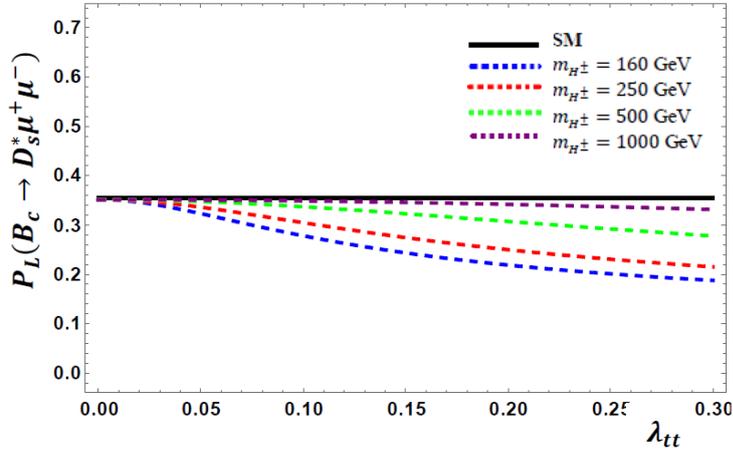

Fig. 8 (c): Variation of polarization fraction ($P_L$) of $B_c \to D_s^* \mu^+ \mu^-$ with $\lambda_{tt}$ in type III 2HDM

From Fig. 3 it is clear that the differential branching ratio for $B_c \to D_s \mu^+ \mu^-$ mostly decreases in all types of 2HDM, but some bounds on model parameters could raise those values. In type I 2HDM (fig. 3 (a)) we have seen that within the range of parameters $2.5 \leq \tan\beta \leq 3$ and $160 < m_{H^\pm} < 250$ GeV the differential branching ratio increases, whereas in type II (fig. 3 (b)) the region of $\tan\beta$ is narrowed down to $2.2 \leq \tan\beta \leq 2.4$ for the noticeable increment. In Fig. 4 it is shown that for $B_c \to D_s^* \mu^+ \mu^-$ the differential branching ratio decreases in type I, increases in type II and type III satisfies the SM value with some increment for different parameter choices. Figs. 5 and 6 depict that increase in $A_{P_L}$ for $B_c \to D_s \mu^+ \mu^-$ channel is very small in 2HDM, but the effect is noticeable for the other mode. From Figs. 7 and 8, it is clear that the observables $A_{FB}$ and $P_L$ are decreased for all three types of 2HDM. The deviation of each observable from the SM with varying model parameter and $m_{H^\pm}$ are shown in the above figures. Now we represent the variation of each parameter with momentum transfer $q^2$.

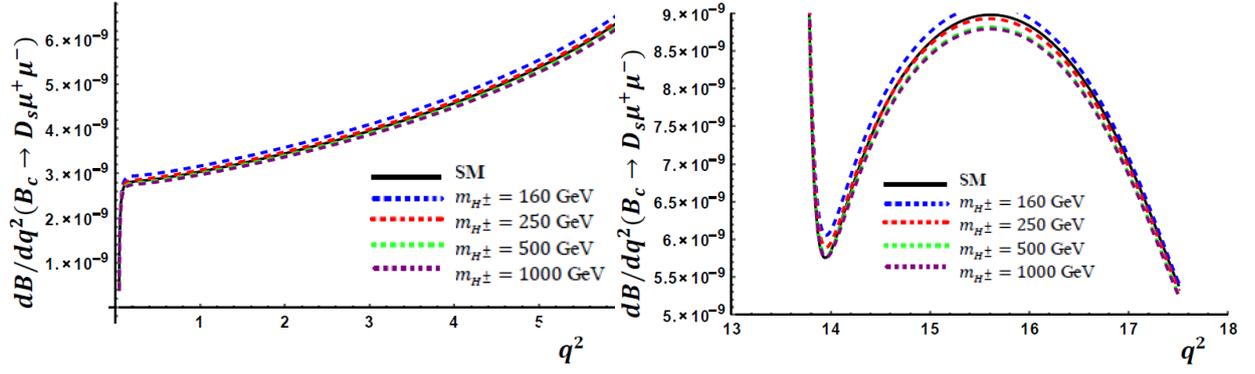

Fig. 9: Variation of differential branching ratio of $B_c \to D_s \mu^+ \mu^-$ with $q^2$ in type III

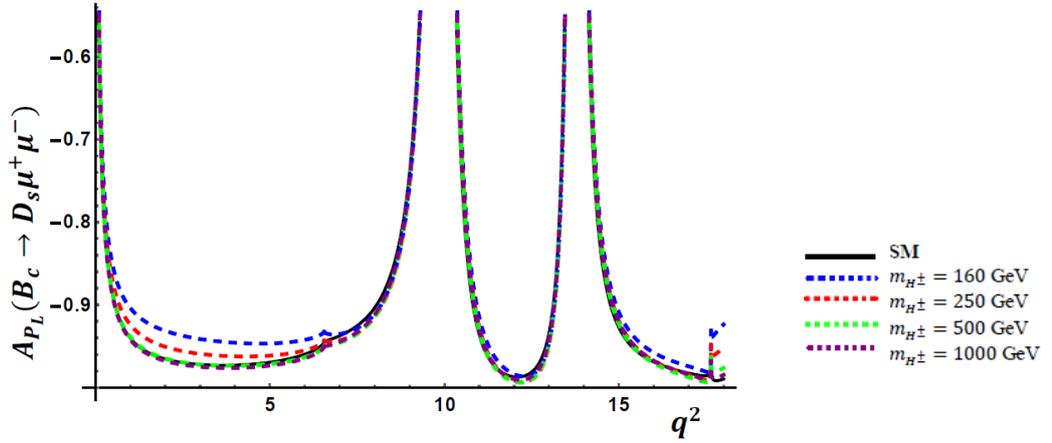

Fig. 10: Variation of lepton polarization asymmetry ($A_{P_L}$) of $B_c \to D_s \mu^+ \mu^-$ with $q^2$ in type III 2HDM

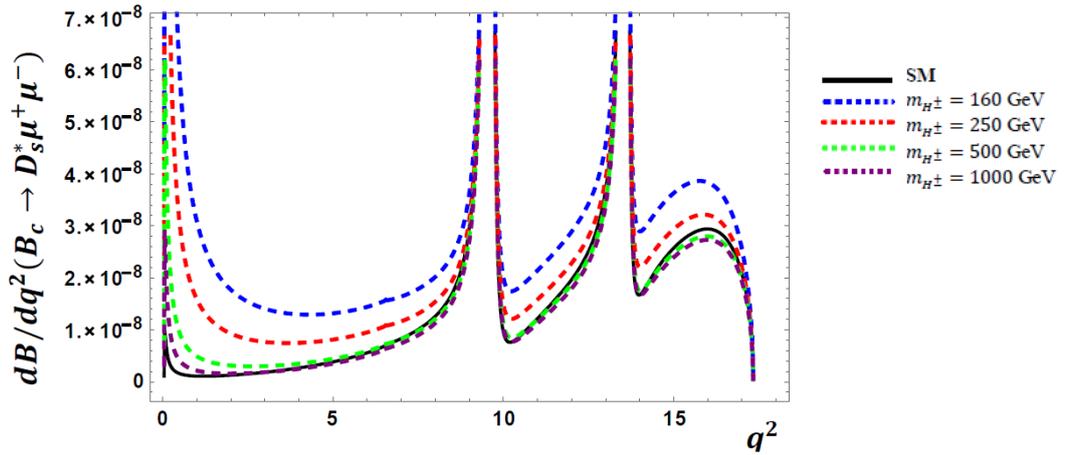

Fig. 11: Variation of differential branching ratio of $B_c \to D_s^* \mu^+ \mu^-$ with $q^2$ in type III

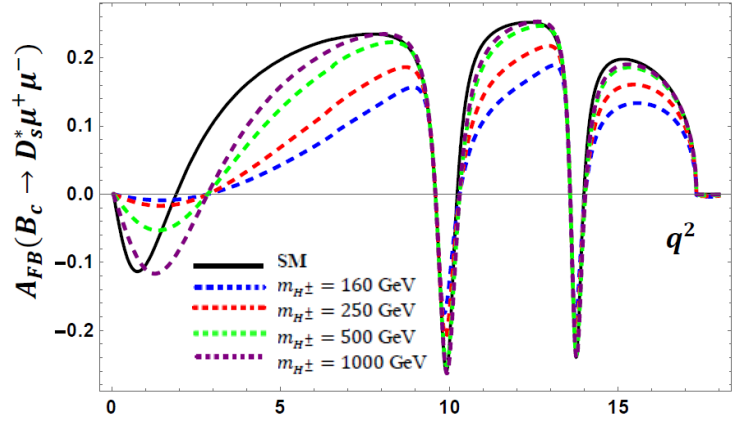

Fig. 12: Variation of forward-backward asymmetry ($A_{FB}$) of $B_c \to D_s^* \mu^+ \mu^-$ with $q^2$ in type III 2HDM

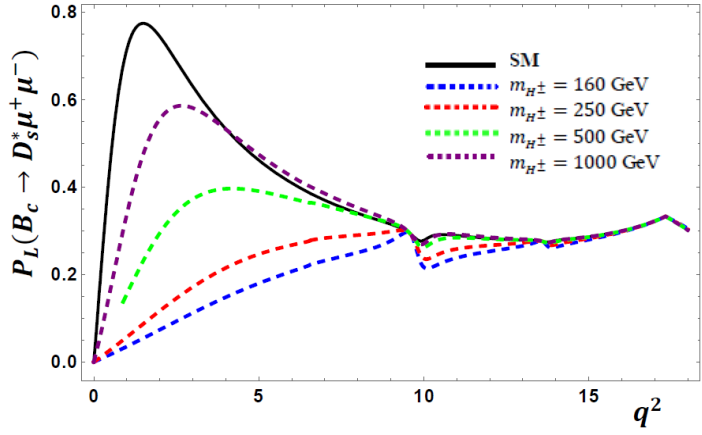

Fig. 13: Variation of polarization fraction ($P_L$) of $B_c \to D_s^* \mu^+ \mu^-$ with $q^2$ in type III 2HDM

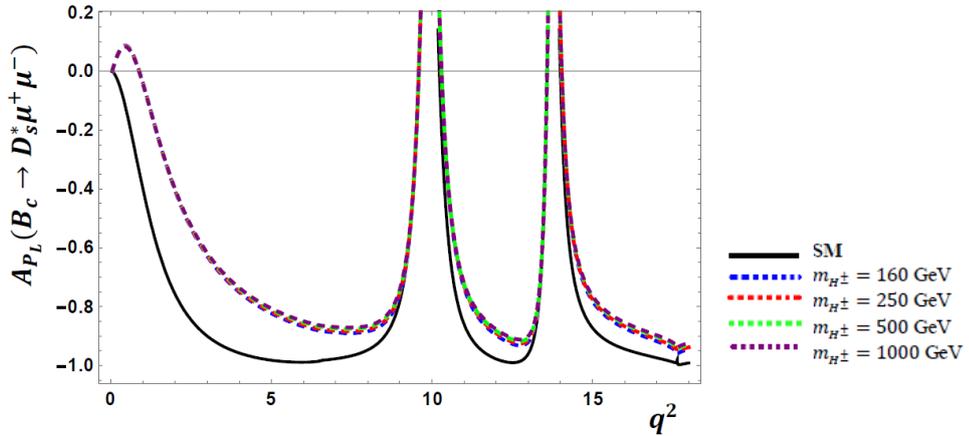

Fig. 14: Variation of lepton polarization asymmetry ($A_{P_L}$) of $B_c \to D_s^* \mu^+ \mu^-$ with $q^2$ in type III 2HDM

Figs. 9 and 11 show the deviations of differential branching ratio for both the decay channels in 2HDM from their SM value. The effect of new charged Higgs boson is not substantial enough for $B_c \to D_s \mu^+ \mu^-$ but significantly large for $B_c \to D_s^* \mu^+ \mu^-$. Here we have shown the differential branching ratio of $B_c \to D_s \mu^+ \mu^-$ for two different regions, i.e., the regions $1 < q^2 < 6$ GeV$^2$ and $13 < q^2 < 18$ GeV$^2$ to understand the slight variations in large scale measurements. For another mode i.e. $B_c \to D_s^* \mu^+ \mu^-$ the maximum deviation arises for $0.15 < \lambda_{tt} < 0.3$ and $160 < m_{H^\pm} < 250$ GeV. Here we have shown the total variation through the whole kinematical region. This channel is more interesting as the branching ratio is increased from SM value in whole $q^2$ region providing a higher possibility of experimental detection of this decay mode. In the resonance region both the SM as well as type III 2HDM curves are suffering from huge uncertainty due to $c\bar{c}$ loops, providing two large infinite peaks. Thus NP and SM could not be distinguished in those resonance peak regions. Also, NP curves and SM curve merge together for larger $q^2$ i.e. $q^2 > 16.5$ GeV$^2$. We have studied all these variations keeping the model parameters fixed at $\lambda_{tt} = 0.3$ and $\lambda_{bb} = 50$.

In the SM, the lepton polarization asymmetry ($A_{P_L}$) for $B_c \to D_s \mu^+ \mu^-$ is almost $\sim -0.7$ for both lower and higher $q^2$ region and $\sim -0.9$ for middle $q^2$ region. The effect of 2HDM to $A_{P_L}$ is quite small for $B_c \to D_s \mu^+ \mu^-$. On the other hand, the maximum value of this parameter for the channel $B_c \to D_s^* \mu^+ \mu^-$ in the SM is $\sim -0.16$, $\sim -0.8$ and $\sim -0.67$ for lower, middle and higher $q^2$ respectively. From Fig. 10 it is very clear that the max value of $A_{P_L}$ could be found for $m_{H^\pm} = 160$ GeV whereas in Fig. 14 this parameter increases in similar manner for all the masses and thus we can state that the mass of charged Higgs is not affecting much to this observable for $B_c \to D_s^* \mu^+ \mu^-$ process. It should be noted that the sign of $A_{P_L}$ of $B_c \to D_s^* \mu^+ \mu^-$ is very sensitive to the charged Higgs contribution and becomes positive for the lower region of $q^2$ where SM could not explain this ambiguity. This property might be thought-out to be one of the most favorable probes in searching for new physics beyond the SM.

Interestingly the forward-backward asymmetry ($A_{FB}$) is found to be present only for $B_c \to D_s^* \mu^+ \mu^-$ channel in this study. Fig. 12 depicts the dependence of forward-backward asymmetry i.e. $A_{FB}$ for the leptons produced in $B_c \to D_s^* \mu^+ \mu^-$ transition on the square of momentum transfer. Here it is found that the zero-crossing point of $A_{FB}$ shifts to $q^2 \sim 2.9$ GeV$^2$ in 2HDM from its SM zero point $q^2 \sim 1.9$ GeV$^2$. Thus determining the sign of $A_{FB}$ in this domain could provide us some clear information about the existence of charged Higgs particle.

For $B_c \to D_s \mu^+ \mu^-$, the longitudinal polarization fraction ($P_L$) has unit value. We observe no variation of this value as $D$ meson does not have any polarization direction. In Fig. 13, the polarization fraction $P_L$ of the vector meson $D^*$ plays quite an interesting role having considerable effects from the charged Higgs boson. The values have been decreased with an increase in $m_{H^\pm}$ within the region $1 < q^2 < 10$ GeV$^2$ significantly. For example, with $m_{H^\pm} = 160$ GeV the values of $P_L$ decreases to about 70% of the SM result. The higher region is not much affected by the presence of new charged Higgs boson. Therefore, it is hoped that measurements of polarization fraction could be a promising tool for establishing the 2HDM.

For the analysis of LFU parameters, we consider only the middle and higher momentum transfer regions as the lower region is not much sensitive in our analysis. Within the SM, the ratios $R_{D_s}$ and $R_{D_s^*} \sim 1$ for both the channels for $1 < q^2 < 6$. In the region $13.6 < q^2 < 17.5$, $R_{D_s}$ and $R_{D_s^*}$ are slightly less than unity. In 2HDM for $B_c \to D_s \mu^+ \mu^-$ channel, $\lambda_{tt} = 0.05$ decreases the ratio whereas $\lambda_{tt} = 0.3$ drives them above the SM. For $B_c \to D_s^* \mu^+ \mu^-$ channel, the value of $R_{D_s^*}$ is increased with $0.05 \leq \lambda_{tt} \leq 0.3$ and $160$ GeV $\leq m_H \leq 1000$ GeV within $1 < q^2 < 6$. But in higher $q^2$ region the value decreases slightly with some other choices of $m_{H^\pm}$.

## VI. Conclusion

In this paper, we have investigated the decay modes $B_c \to D_s \mu^+ \mu^-$ and $B_c \to D_s^* \mu^+ \mu^-$ in the 2HDM. We have analyzed various decay observables such as branching ratios, forward-backward asymmetry, polarization fraction, lepton polarization asymmetry and LFU parameters using a relativistic quark model. We have found that the branching ratio for pseudoscalar mode of $B_c$ meson (i.e. $B_c \to D_s \mu^+ \mu^-$) is less sensitive to the new charged Higgs particle than its vector mode (i.e. $B_c \to D_s^* \mu^+ \mu^-$). Branching ratio for the vector mode consists of some considerable increment in 2HDM than its SM predictions and decreasing the value of $m_{H^\pm}$ increases the branching ratio. The forward-backward asymmetry carries out much attention in the lower momentum transfer region. Moreover, in 2HDM type III, zero point of the SM shifts towards higher $q^2$ and the value of $A_{FB}$ in lower $q^2$ increases with decreasing value of $m_{H^\pm}$. This indicates that $m_{H^\pm} = 160$ GeV provides maximum value of $A_{FB}$ within $0.045 < q^2 < 3$ GeV$^2$. The lepton polarization asymmetry ($A_{P_L}$) deviates sizably from that of the SM and especially for $B_c \to D_s^* \mu^+ \mu^-$ the parameter $A_{P_L}$ gets positive value in lower $q^2$ which is a fascinating fact for establishing new physics. The polarization fraction $P_L$ is also sensitive to 2HDM with notable decrement from the SM results with decreasing $m_{H^\pm}$. The analysis of LFU ratios $R_{D_s}$ and $R_{D_s^*}$ are quite exciting in our study because there are both increment and decrement from the SM values for different choices of model parameters and $m_{H^\pm}$. So further experimental verification is needed to determine the suitability of our model. Then only any strict explanation of lepton flavor universality violation from these two ratios in 2HDM could provide the limits to the Yukawa couplings and mass of charged Higgs particle.

Concluding the whole study, it might be stated that the semileptonic channels of rare charm B meson e.g. $B_c \to D_s \mu^+ \mu^-$ and $B_c \to D_s^* \mu^+ \mu^-$ are not being experimentally observed so far. The sensitive measurements of different decay observables of these channels are needed to open up new possibilities towards new physics beyond the SM. We hope our study for these decay modes in 2HDM will play a crucial role in future.

### Acknowledgment

We thank the reviewer for valuable comments which improve quality of our manuscript. Maji is thankful to DST, Govt. of India for providing INSPIRE Fellowship (IF160115). Nayek and Sahoo are grateful to SERB, DST, Govt. of India for financial support (EMR/2015/000817). Biswas thanks NIT Durgapur for providing fellowship.

### References:

[1] R. Aaij et al. (LHCb Collaboration), Phys. Rev. Lett. **111** (2013) 191801, [arXiv: 1308.1707].

[2] R. Aaij et al. (LHCb Collaboration), J. High Energy Phys. **07** (2013) 084, [arXiv: 1305.2168].

[3] R. Aaij et al. (LHCb Collaboration), Phys. Rev. Lett. **113** (2014) 151601, [arXiv: 1406.6482]; Phys. Rev. Lett. **122** (2019) 191801; A. Abdesselam et al. (Belle Collaboration), arXiv: 1908.01848 (2019).

[4] R. Aaij et al. (LHCb Collaboration), J. High Energy Phys. **08** (2017) 055, arXiv: 1705.05802; A. Abdesselam et al. (Belle Collaboration), arXiv: 1904.02440 (2019).

[5] A. K. Alok et al., J. High Energy Phys. **09** (2018) 152, arXiv: 1710.04127.

[6] U. Egede, T. Hurth, J. Matias, M. Ramon, and W. Reece, J. High Energy Phys. **11** (2008) 032, arXiv: 1005.0571.

[7] W. Altmannshofer, A. J. Buras, D. M. Starub, and M. Wick, J. High Energy Phys. **04** (2009) 022.

[8] G. Buchalla, G. Hiller, and G. Isidori, Phys. Rev. D **63** (2000) 014015.

[9] C. Bird, P. Jackson, R. Kowalewski, and M. Pospelov, Phys. Rev. Lett. **93** (2004) 201803.

[10] S. S. Gershtein, V. V. Kiselev, A. K. Likhoded, and A. V. Tkabladze, Usp. Fiz. Nauk **165** (1995) 3 [Phys. Usp. **38** (1995) 1].

[11] R. Dhir, and R. C. Verma, Phys. Scr. **82** (2010) 065101.

[12] A. Faessler, T. Gutsche, M. A. Ivanov, J. G. Körner, and V. E. Lyubovitskij, Eur. Phys. J. direct **4** (2002) 1.

[13] C. Q. Geng, C. W. Hwang, and C. C. Liu, Phys. Rev. D **65** (2002) 094037.

[14] D. Ebert, R. N. Faustov, and V. O. Galkin, Phys. Rev. D **82** (2010) 034032.

[15] I. Ahmed, M. A. Paracha, M. Junaid, A. Ahmed, A. Rehman, and M. J. Aslam, arXiv: 1107.5694.

[16] A. Ahmed, I. Ahmed, M. A. Paracha, M. Junaid, A. Rehman, and M. J. Aslam, arXiv: 1108.1058.

[17] K. Azizi, F. Falahati, V. Bashiry, and S. M. Zebarjad, Phys. Rev. D **77** (2008) 114024.

[18] K. Azizi, and R. Khosravi, Phys. Rev. D **78** (2008) 036005

[19] C. Q. Cheng, C. W. Hwang, and C. C. Liu, Phys. Rev. D **65** (2002) 094037.

[20] H. M. Choi, Phys. Rev. D **81** (2010) 054003.

[21] W. F. Wang, X. Yu, C. D. Lu, and Z. J Xiao, Phys. Rev. D **90** (2014) 094018.

[22] L. Lin-Xia, Z. Guo-Fang, and W. Shuai-Wei, Commun. Theor. Phys. **59** (2013) 187.

[23] U. O. Yilmaz, arXiv: 1204.1261.

[24] U. O. Yilmaz, and G. Turan, Eur. Phys. J. C **51** (2007) 63.

[25] R. Dutta, Phys. Rev. D **100** (2019) 075025, arXiv: 1906.02412.

[26] D. Atwood et al., Phys. Rev. D **55** (1997) 3156, arXiv: hep-ph/9609279 [JL-TH-96-15, JLAB-THY-96-08, and CEBAF-PREPRINT-JL-TH-96-15].

[27] G. C. Branco et al., Phys. Rept. **516** (2012) 1, arXiv: 1106.0034.

[28] F. Kling et al., J. High Energy Phys. **09** (2016) 093, arXiv: 1604.01406 [FERMILAB-PUB-16-093-T].

[29] T. Barakat, J. Phys. G **24** (1998) 1903.

[30] T. Barakat, Il Nuovo Cimento **112** (1999) 697.

[31] T. M. Aliev and E. O. Iltan, Phys. Rev. D **58** (1998) 095014.

[32] T. M. Aliev and E. O. Iltan, J. Phys. G **25** (1999) 989.

[33] A. K. Grant, Phys. Rev. D **51** (1995) 207.

[34] J. Kalinowski, Phys. Lett. B **245** (1990) 201.

[35] D. Chowdhury, and O. Eberhardt, J. High Energy Phys. **05** (2018) 161.

[36] J. Lorenzo Diaz-Cruz, Rev. Mex. Fis. **65** (2019) 419.

[37] P. Arnan, D. Becirevic, F. Mescia, and O. Sumensari, Eur. Phys. J. *C* **77** (2017) 796

[38] J. F. Gunion, and H. E. Haber, Phys. Rev. D **67** (2003) 075019, arXiv: hep-ph/0207010.

[39] A. Barroso, P. M. Ferreira, I. P. Ivanov, and R. Santos, J. High Energy Phys. **06** (2013) 045, arXiv: 1303.5098.

[40] S. Kanemura, T. Kubota, E. Takasugi, Phys. Lett. B **313** (1993) 155, arXiv: hep-ph/9303263.

[41] B. Swiezewska, Phys. Rev. D **88** (2013) 055027 (Erratum: Phys. Rev. D **88** (2013) 119903), arXiv: 1209.5725.

[42] J. Haller et al., Eur. Phys. J. C **78** (2018) 675, arXiv: 1803.01853.

[43] G. Cvetic, C. S. Kim, S. S. Hwang, Int. J. Mod. Phys. A **14** (1999) 769, arXiv: hep-ph/9706323.

[44] T.P. Cheng and M. Sher, Phys. Rev. D **35** (1987) 3484.

[45] C. S. Kim, Y. W. Yoon, and X. Yuan, J. High Energy Phys. **12** (2015) 038.

[46] G. C. Branco et al., Phys. Rep. **516** (2012) 1.

[47] E. Golowich, J. Hewett, S. Pakvasa, and A. A. Petrov, Phys. Rev. D **76** (2007) 095009, arXiv: 0705.3650.

[48] S. L. Glashow, and S. Weinberg, Phys. Rev. D **15** (1977) 1958.

[49] T. M. Aliev, and E. O. Iltan, J. Phys. G: Nucl. Part. Phys. **25** (1999) 989, arXiv: hep-ph/9803272.

[50] T. Enomoto, and R. Watanabe, J. High Energy Phys. **05** (2016) 002.

[51] F. Falahati, and R. Khosravi, Phys. Rev. D **85** (2012) 075008.

[52] R. F. Alnahdi, T. Barakat, and H. A. Alhendi, Prog. Theor. Exp. Phys. **2017** (2017) 073B04.

[53] B. Grinstein, M. J. Savage, and M. B. Wise, Nucl. Phys. B **319** (1989) 271.

[54] Y. B. Dai, C. S. Huang, J. T. Li, and W. J. Li, Phys. Rev. D **67** (2003) 096007.

[55] D. Melikhov, N. Nikitin, and S. Simula, Phy. Lett. B **430** (1998) 332.

[56] A. J. Buras, M. Misiak, M. Muenz, and S. Pokorski, Nucl. Phys. B **424** (1994) 374, arXiv: hep-ph/9311345

[57] G. Erkol, and G. Turan, Eur. Phys. J. C **25** (2002) 575.

[58] M. Misiak, Nucl. Phys. B **393**, 23 (1993); Erratum ibid B **439** (1995) 161.

[59] A. J. Buras, and M. Muenz, Phys. Rev. D **52** (1995) 186, arXiv: hep-ph/950128.

[60] N. Katirci, and K. Azizi, J. Phy. G: Nucl. Part. Phys. **40** (2013) 085005.

[61] J. G. Korner, and G. A. Schuler, Z. Phys. C **46** (1990) 93.

[62] A. Kadeer, J. G. Korner, and U. Moosbrugger, Eur. Phys. J. C **59** (2009) 27.

[63] W. L. Ju, G. L. Wang, H. F. Fu, T. H. Wang, and Y. Jiang, J. High Energy Phys. **04** (2014) 065, arXiv: 1307.5492.

[64] M. Tanabashi et al., Phys. Rev. D **98** (2018) 030001.

[65] G. Abbiendi et al. [ALEPH, DELPHI, L3, OPAL, LEP Collaboration], Eur. Phys. J. C **73** (2013) 2463, arXiv: 1301.6065.

[66] G. Aad et al. [ATLAS Collaboration], J. High Energy Phys. **03** (2015) 088, arXiv: 1412.6663.

[67] V. Khachatryan et al. [CMS Collaboration], J. High Energy Phys. **11** (2015) 018, arXiv: 1508.07774.

[68] G. Aad et al. [ATLAS Collaboration], Eur. Phys. J. C **73** (2013) 2465, arXiv: 1302.3694.

[69] V. Khachatryan et al. [CMS Collaboration], J. High Energy Phys. **12** (2015) 178, arXiv: 1510.04252.

[70] CMS Collaboration (2016) [CMS-PAS-HIG-16-030].
[71] M. Aaboud et al. [ATLAS Collaboration], Phys. Lett. B **759** (2016) 555, arXiv: 1603.09203.
[72] ATLAS Collaboration (2016) [ATLAS-CONF-2016-088].
[73] CMS Collaboration (2016) [CMS-PAS-HIG-16-031].
[74] ATLAS Collaboration (2016) [ATLAS-CONF-2016-089].
[75] CMS Collaboration, J. High Energy Phys. **07** (2019) 142, arXiv: 1903.04560.
[76] ATLAS Collaboration, J. High Energy Phys. **11** (2018) 085, arXiv: 1808.03599.
[77] A. Arbey, F. Mahmoudi, O. Stal, and T. Stefaniak, Eur. Phys. J. C **78** (2018) 182.
[78] Y. Liu, L. L. Liu, and X. H. Guo, arXiv:1503.06907
[79] D. Atwood, L. Reina, and A. Soni, Phys. Rev. D **55** (1997) 3156.
[80] D. Browser-Chao, K. Cheung, and W. Y. Keung, Phys. Rev. D **68** (1999) 115006.
[81] C. S. Huang, and S. H. Zhu, Phys. Rev. D **68** (2003) 114020.
[82] M. S. Alam et al., CLEO Collaboration, in ICHEP 98 Conference, 1998.
[83] Basabendu Barman, Debasish Borah, Lopamudra Mukherjee, and Soumitra Nandi, Phys. Rev. D 100 (2019) 115010, arXiv: 1808.06639.
[84] Talk by R. Briere, Proceedings of ICHEP98, Vancouver, Canada, 1998, Report /no. CLEO-CONF-98-17, ICHEP98-1011.
[85] R. Barate et al. (ALEPH Collaboration), Phys. Lett. B **429** (1998) 169.
[86] K. Kiers, and A. Soni, Phys. Rev. D **56** (1997) 5786.
[87] A. Stahl, and H. Voss, Z. Phys. C **74** (1997) 73.
[88] M. Hussain et al., Phys. Rev. D **95** (2017) 075009.